\documentclass[journal]{IEEEtran}

\ifCLASSINFOpdf
\usepackage[pdftex]{graphicx}
\DeclareGraphicsExtensions{.pdf,.jpeg,.png}
\else
\usepackage[dvips]{graphicx}
\fi
\usepackage{graphicx}
\usepackage{epstopdf}
\usepackage[cmex10]{amsmath}
\usepackage{amssymb}
\usepackage{wasysym}
\usepackage{algorithm}
\usepackage{algorithmic}
\usepackage{array}
\usepackage{cite}
\usepackage{color}
\usepackage{url}
\usepackage{bm}
\usepackage{enumerate}
\usepackage{latexsym}
\usepackage{epsfig}
\usepackage{cite}
\usepackage{caption}
\usepackage{subfigure}
\let\labelindent\relax
\usepackage{enumitem}

\usepackage[colorlinks,linkcolor=red,anchorcolor=red,citecolor=green]{hyperref}

\begin{document}

\title{Beam Switching Based Beam Design for High-Speed Train mmWave Communications}

\author{Jingjia~Huang,~\IEEEmembership{Graduate~Student~Member,~IEEE}, Chenhao~Qi,~\IEEEmembership{Senior~Member,~IEEE}, \\ Octavia~A.~Dobre,~\IEEEmembership{Fellow,~IEEE}, and Geoffrey Ye Li,~\IEEEmembership{Fellow,~IEEE}
	\thanks{This work was supported in part by the National Natural Science Foundation of China under Grants 62071116 and U22B2007. (\textit{Corresponding author: Chenhao Qi}.)}
	\thanks{Jingjia~Huang and Chenhao~Qi are with the School of Information Science and Engineering, Southeast University, Nanjing 210096, China (e-mail:~\{jiah,~qch\}@seu.edu.cn).}
	\thanks{Octavia~A.~Dobre is with the Faculty of Engineering and Applied Science, Memorial University, St. John’s, NL A1C 5S7, Canada (e-mail: odobre@mun.ca).}
	\thanks{Geoffrey~Ye~Li is with the Department of Electrical and Electronic Engineering, Imperial College London, SW7 2AZ London, U.K. (e-mail: geoffrey.li@imperial.ac.uk).}
}
\markboth{Accepted by IEEE Transactions on Wireless Communications}
{}


\maketitle




\begin{abstract}
For high-speed train (HST) millimeter wave (mmWave) communications, the use of narrow beams with small beam coverage needs frequent beam switching, while wider beams with small beam gain leads to weaker mmWave signal strength. In this paper, we consider beam switching based beam design, which is formulated as an optimization problem aiming to minimize the number of switched beams within a predetermined railway range subject to that the receiving signal-to-noise ratio (RSNR) at the HST is no lower than a predetermined threshold. To solve this problem, we propose two sequential beam design schemes, both including two alternately-performed stages. In the first stage, given an updated beam coverage according to the railway range, we transform the problem into a feasibility problem and further convert it into a min-max optimization problem by relaxing the RSNR constraints into a penalty of the objective function. In the second stage, we evaluate the feasibility of the beamformer obtained from solving the min-max problem and determine the beam coverage accordingly. Simulation results show that compared to the first scheme, the second scheme can achieve 96.20\% reduction in computational complexity at the cost of only 0.0657\% performance degradation.
\end{abstract}

\section{Introduction}\label{Introduction}
\IEEEPARstart{F}{or} high-speed train (HST) millimeter wave (mmWave) communications, one of the main challenges is the fast time-varying fading caused by the high mobility of the HST~\cite{he5GRailwaysNext2022}. The short channel coherence time restricts the time available for conventional beam management procedures, such as beam training and tracking~\cite{10288339}. As an alternative, beam switching proves more suitable in this scenario, due to its simpler mechanism and lower overhead~\cite{cuiOptimalNonuniformSteady2018,vaBeamSwitching2015}. Specifically, each beam used in the beam switching has a predesigned coverage for the railway range. When communicating with the HST, the base station (BS) achieves beam alignment through simply switching these predesigned beams according to the location of the HST. Note that the location of the HST is currently available since it is obtained from the railway control system to ensure the railway security~\cite{maTrainPositionBased2017}. 

For HST mmWave communications, the procedures to predesign beams for beam switching differ from conventional beamforming optimized based on instantaneous channel state information (CSI)~\cite{gaoEfficientHybrid2020,liThroughputMaximization2023,chenJointDesign2023,9726800}. Predesigning beams involves optimizing the beam coverage based on statistical CSI and geometric information of the railway. By assuming a uniform beam width (UBW) for all the predesigned beams, the beam coverage is optimized to maximize the average data rate~\cite{vaBeamSwitching2015}. However, this assumption leads to considerable variations in the receiving signal-to-noise ratio (RSNR) at different HST locations, since the path loss varies as the HST moves but the level of beam gain remains constant. To address this issue, a non-uniform beam width (NUBW) optimization scheme is proposed, where the coverage of each beam is optimized to minimize the data rate gap among beams~\cite{cuiOptimalNonuniformSteady2018}. The increased degree of freedom introduced by adding optimization variables enables NUBW to achieve better RSNR stability than UBW. To consider the beamformer design subject to hardware constraints that is not included in~\cite{cuiOptimalNonuniformSteady2018}, an analog precoding codebook, where wide beams are designed by switching off some antennas, is developed with closed-form solutions based on the beam coverage designed by an equal spacing coverage (ESC) scheme~\cite{liuAdaptiveNonUniform2023a}. With each beam covering the same length of railway range, ESC can approximate the average data rate performance of NUBW. Note that in all the aforementioned studies, the beam gain used in the beam coverage optimization is not generated by actual beamformers, but approximated by simple functions of the beam width. Although beamformers can be designed separately after the beam coverage optimization, the mismatch between the approximated and actual beam gain may lead to performance loss, such as unstable RSNR. 

Most existing studies on predesigning beams for HST mmWave communications optimize the beam coverage with a given number of switched beams, whereas the problem of minimizing the number of switched beam receives limited attention. To minimize the number of switched beams, the beam coverage is optimized without considerations of the beamformer design  in~\cite{yanStableBeamformingLow2018}. Note that avoiding frequent beam switching through predesigning beams is essential for the performance of HST mmWave communications. Specifically, if narrow beams are used by the BS, the duration that the HST falls within the same beam coverage can be less than one millisecond in certain cases~\cite{guanMillimeterWave2017}. Switching beams within such a short duration may result in beam misalignment. To avoid frequent beam switching, wide beam can be considered, where the beam gain should be enough to guarantee the quality of service (QoS).

In this paper, we consider beam switching based beam design by jointly optimizing the beam coverage and beam gain in a downlink HST mmWave system. Our main contributions are summarized as follows.
\begin{itemize}
	\item Since the HST may be very close to the BS, we establish a near-field channel model to characterize the time-varying channel between the BS and the HST. Then, we formulate the optimization problem aiming to minimize the number of switched beams within a predetermined railway range, subject to constant modulus constraints and RSNR constraints that require the RSNR at each HST location no lower than a predetermined threshold. Due to a large number of non-convex constraints and the interdependent optimization variables, the original problem is difficult to solve. We propose a sequential beam design approach to convert the problem into a series of beam design sub-problems that aim at maximizing the beam coverage subject to the RSNR constraints and constant modulus constraints.
	\item For each sub-problem, we propose a framework including two stages (TSs) that are performed alternately. In the first stage, given a fixed beam coverage, we transform the sub-problem into a feasibility problem and further convert it into a non-convex min-max optimization problem by relaxing the RSNR constraints into a penalty of the objective function. In the second stage, we evaluate the feasibility of the beamformer obtained from solving the min-max problem and determine the beam coverage accordingly. 
	\item Under the TS framework, we propose two schemes to solve each sub-problem. For the first scheme, the min-max problem is solved based on semidefinite relaxation (SDR) and a difference-of-convex (DC) algorithm, while the beam coverage is determined by a bisection search (BiS) method. For the second scheme, the min-max problem is solved based on a proximal-point (PP) method and a primal-dual gradient (PDG) algorithm with closed-form solutions, while the beam coverage is determined by a mixed search (MS) method.
\end{itemize}

The rest of this paper is organized as follows. The system model is introduced in Section~\ref{SystemModel}. In Section~\ref{ProblemFormulation}, we formulate the optimization problem and introduce the sequential beam design approach. Section~\ref{Method} and Section~\ref{Method1} present the two schemes. Simulation results and relevant analysis are provided in Section~\ref{Simulation}. Finally, the conclusion is presented in Section~\ref{Conclusion}. 

The notations are defined as follows. Symbols for matrices (upper case) and vectors (lower case) are in boldface. $(\cdot)^{\text{T}}$, $(\cdot)^{\text{H}}$, $\textrm{Tr}\{\cdot\}$ and $\vert\cdot\vert$ denote the transpose, conjugate transpose (Hermitian), trace and absolute value, respectively. $\Vert\boldsymbol{a}\Vert_2$ denotes the $\ell_2$-norm of a vector $\boldsymbol{a}$, while $\Vert\boldsymbol{A}\Vert_2$ denotes the spectral norm of a matrix $\boldsymbol{A}$. $\Vert\boldsymbol{A}\Vert_{*}$ denotes the nuclear norm of a matrix $\boldsymbol{A}$. $\textrm{Diag}\{\boldsymbol{a}\}$ denotes a diagonal matrix with its diagonal entries represented by a vector $\boldsymbol{a}$. $\mathbb{C}$, $\mathbb{R}$, $\mathcal{O}$ and $\mathcal{N}$ denote the set of complex number, set of real number, order of complexity and Gaussian distribution, respectively. $\mathrm{Re}\{\cdot\}$ and $\mathrm{Im}\{\cdot\}$ denote the real part and imaginary part of a complex number, respectively. The $m$th entry of a vector $\boldsymbol{a}$ is denoted as $[\boldsymbol{a}]_{m}$. The entry at the $m$th row and $n$th column of a matrix $\boldsymbol{A}$ is denoted as $[\boldsymbol{A}]_{m,n}$.

\section{System Model}\label{SystemModel}
\begin{figure}[!t]
	\centering
	\includegraphics[width=0.58\linewidth]{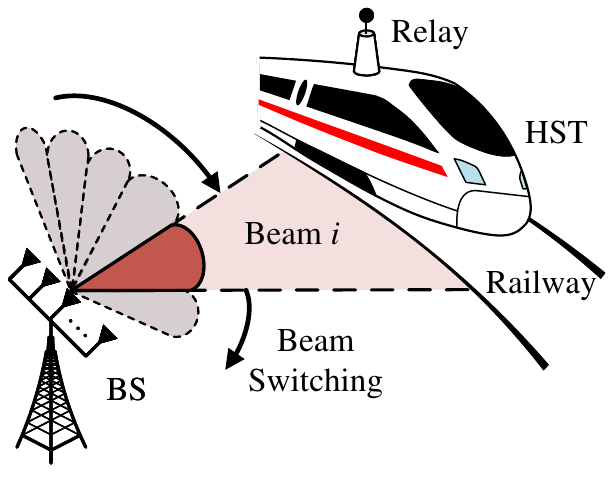}
	\vspace{-2mm}
	\caption{Illustration of HST mmWave communications.}
	\vspace{-3mm}
	\label{FigSystemModel}
\end{figure}

\begin{figure}[!t]
	\centering
	\includegraphics[width=0.78\linewidth]{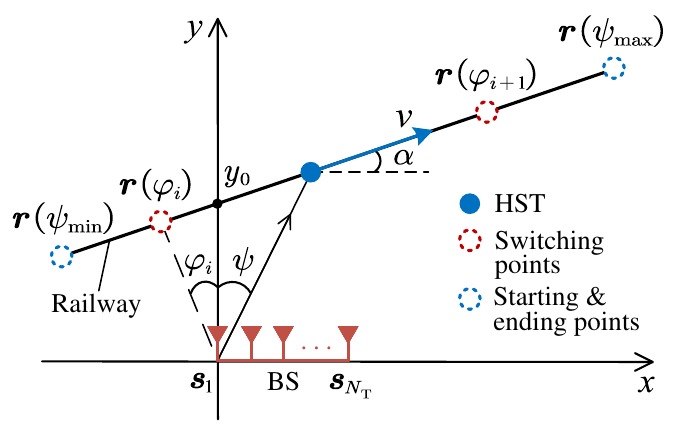}
	\vspace{-2mm}
	\caption{Geometric illustration of the considered system.}
	\vspace{-3mm}
	\label{FigSystemModelGeometry}
\end{figure}

As shown in Fig.~\ref{FigSystemModel}, we consider a downlink HST mmWave system comprising a BS and an HST. The BS is equipped with a uniform linear array (ULA) consisting of $N_\mathrm{T}$ antennas, where the antenna spacing is denoted as $\delta_{\mathrm{T}}$. The direct transmission of mmWave signals from the BS to user terminals (UTs) within the HST may encounter severe penetration loss~\cite{liMobilitySupportMillimeter2022}. To address this issue, adopting a two-hop architecture has been regarded as a suitable strategy~\cite{fan20165g}. In light of this, we assume that an antenna array is equipped on top of the HST, relaying mmWave signals from the BS to UTs within the HST. To focus on the transmit beamforming at the BS, we further assume that the relay receives signals omnidirectionally. 

\subsection{Geometric System Model}\label{GeoModel}
To give a geometric system model, we set up a two-dimensional coordinate system as shown in Fig.~\ref{FigSystemModelGeometry}. The ULA at the BS is oriented horizontally, with the coordinate of each antenna denoted as $\boldsymbol{s}_{n}=\big[(n-1)\delta_{\mathrm{T}},0\big]^{\mathrm{T}},\ n\in\mathcal{I}_{\mathrm{N}}\triangleq\{1,2,\dots,N_{\mathrm{T}}\}$. The coordinate of the relay on the HST can be expressed based on the geometric information of the railway as
\begin{equation}\label{AoD2r}
	\boldsymbol{r}(\psi)=\bigg[\frac{y_{0}\tan\psi}{1-\tan\alpha\tan\psi},\frac{y_{0}}{1-\tan\alpha\tan\psi}\bigg]^{\mathrm{T}},
\end{equation}
where $\psi\in\big(-\pi/2,\pi/2-\alpha\big)$ is the angle of departure (AoD), $\alpha$ is the angle between the railway and the $x$-axis, and $y_0$ is the intersection point of the railway and the $y$-axis. Without loss of generality, we set $\alpha\ge0$ and $y_{0}>0$.

As the HST moves along the railway, the BS performs beam switching based on the estimation of $\boldsymbol{r}(\psi)$ to avoid beam misalignment. Each beam transmitted by the BS is denoted as $\boldsymbol{f}_i \in \mathbb{C}^{N_\mathrm{T}},\ i=1,2,\dots,N$. Since the analog beamforming for mmWave communications is commonly implemented using phase shifting networks, each entry of $\boldsymbol{f}_i$ should be subject to the constant modulus constraint from the phase shifters~\cite{chen2019two}, i.e.,
\begin{equation}\label{Constant_Constraint}
	\big\vert[\boldsymbol{f}_i]_{n}\big\vert=\frac{1}{\sqrt{N_{\mathrm{T}}}},\ \forall n\in \mathcal{I}_{\mathrm{N}}.
\end{equation}
Once the HST is estimated to reach a specific location $\boldsymbol{r}(\varphi_{i+1})$, the BS switches the beam from $\boldsymbol{f}_{i}$ to $\boldsymbol{f}_{i+1}$ in order to maintain a reliable connection, where $\varphi_{i+1}$ is the AoD corresponding to the switching location. In this paper, we aim at minimizing the number of switched beams, represented by $N$, within a given railway range  from $\boldsymbol{r}(\psi_{\min})$ to $\boldsymbol{r}(\psi_{\max})$. Without loss of generality, we require $\varphi_{1}=\psi_{\min}$ and $\varphi_{N+1}\geq\psi_{\max}$.

\subsection{Channel Model}
For HST mmWave communications, the channel coherence time is typically very short due to the high mobility of the HST. For example, in the tunnel scenario with a carrier frequency of 30 GHz, the channel coherence time can be as short as 0.47 ms~\cite{guanRealisticHighSpeedTrain2018}. In such a limited timeframe, acquiring the instantaneous CSI and designing beamforming based on the acquired CSI is challenging. As a more practical alternative, beam switching based beam design is considered in this paper. To support the beam design, we establish a channel model based on geometric and statistical information. The framework, key technologies and application scenarios are included in a pioneer work on railway dedicated communications~\cite{he5GRailwaysNext2022}. 

Suppose the line-of-sight path dominates the signal transmission. The time-varying channel impulse response between the $n$th BS antenna and the HST can be formulated as
\begin{equation}\label{GroundTruthChannelModel}
	h_{n}(\tau,t) =\beta_{n}(t) \exp \big(-j2\pi f_{\mathrm{c}}\tau_{n}(t)\big) \delta\big(\tau-\tau_{n}(t)\big),
\end{equation}
where $\beta_{n}(t)$, $\tau_{n}(t)$, $f_{\mathrm{c}}$ and $\delta(t)$ represent the large-scale fading, propagation delay, carrier frequency and impulse function, respectively. For simplicity, the HST is assumed to move at a constant speed of $v$ m/s within a given time interval $[0,T]$. Then, the AoD at $t\in[0,T]$ can be expressed as
\begin{equation}\label{Psi_t}
	\psi(t) =\arctan\Big(\frac{[\boldsymbol{r}_{1}]_{1}+vt\cos\alpha}{[\boldsymbol{r}_{1}]_{2}+vt\sin\alpha}\Big),
\end{equation}
where $\boldsymbol{r}_{1}$ is a predetermined relay coordinate obtained from the railway control system. Note that variations in velocity and different railway geometries will change the expressions for the AoD and the coordinate of the relay. However, discussions on the system model with arbitrary velocities and railway geometries are reserved for future work due to page limits. Based on $\psi(t)$, the propagation delay can be formulated as 
\begin{equation}
	\tau_{n}(t)= \big\Vert\boldsymbol{r}\big(\psi(t)\big)-\boldsymbol{s}_{n}\big\Vert_2 /c,
\end{equation}
where $c$ is the speed of light. Since $\tau_{n}(t)$ is highly non-linear and its continuity leads to complex semi-infinite programming problems, it is difficult to design beams directly based on~\eqref{GroundTruthChannelModel}. Therefore, to facilitate the subsequent beam design while ensuring the accuracy of the channel model, we divide $[0,T]$ into $M-1$ sub-intervals and approximate $\tau_{n}(t)$ within each sub-interval. The $m$th sub-interval is denoted as $[t_{m},t_{m+1})$, where $\ m=1,2,\dots,M-1$, $t_{1}=0$ and $t_{M}=T$. Within $[t_{m},t_{m+1})$, $\tau_n(t)$ is approximated using the second-order Taylor expansion~\cite{gongDataAidedDopplerCompensation2021,liuNearFieldCommunications2023,10563980}. Define $\psi_{m}\triangleq \psi(t_m)$ and 
\begin{equation}\label{DistanceM}
	r_{m}\triangleq  \Vert\boldsymbol{r}(\psi_{m})-\boldsymbol{s}_{1}\Vert_{2}=\frac{y_{0}\cos\alpha}{\cos(\psi_{m}+\alpha)},
\end{equation} 
for $m\in\mathcal{I}_{\mathrm{M}}\triangleq\{1,2,\dots,M\}$. Then we can approximate $\tau_{n}(t)$ within $[t_{m},t_{m+1})$ by 
\begin{equation}\label{DistanceSimp1}
	\begin{aligned}
	\tau_{n}(t) &\approx \frac{r_{m}}{c}+\frac{v}{c}(t-t_m)\sin(\alpha+\psi_{m}) \\
	&\quad\quad\quad-(n-1)\frac{\delta_{\mathrm{T}}}{c}\sin\psi_{m} + \Delta\tau_n(t),
	\end{aligned}
\end{equation} 
where $\Delta \tau_n(t)$ represents the nonlinear component and can be expressed as
\begin{equation}\label{DistanceSimp2}
	\Delta \tau_n(t)=\frac{\big(v(t-t_{m})\cos(\alpha+\psi_m) -(n-1)\delta_{\mathrm{T}}\cos\psi_{m}\big)^2}{2r_{m}c}.
\end{equation}
Generally, the phase shift introduced by $\Delta\tau_n(t)$ has an upper bound, i.e.,
\begin{equation}
	\big\vert 2\pi(f_{\mathrm{c}}+f)\Delta\tau_n(t)\big\vert\!\leq\! \pi\Big(1\!+\!\frac{B_{\mathrm{f}}}{2f_{\mathrm{c}}}\Big)\frac{v^2(t_{m+1}\!-\!t_{m})^2\!+\!N_{\mathrm{T}}^2\delta_{\mathrm{T}}^2}{r_{m}\lambda_{\mathrm{c}}},
\end{equation}
where $\lambda_{\mathrm{c}}=c/f_{\mathrm{c}}$ is the signal wavelength and $B_{\mathrm{f}}$ is the bandwidth. It can be observed that if $t_{m+1}-t_{m}$ is sufficiently small, i.e., 
\begin{equation}\label{DistanceSimp3Con}
	t_{m+1}-t_{m}\ll \Delta_{m}\triangleq \frac{1}{v}\sqrt{\frac{2r_{m}\lambda_{\mathrm{c}}}{1+B_{\mathrm{f}}/(2f_{\mathrm{c}})}}, 
\end{equation}
the phase shift introduced by the time-varying component in~\eqref{DistanceSimp2} becomes much smaller than $2\pi$, and therefore can be neglected, leading to the approximation
\begin{equation}\label{DistanceSimp3}
	\begin{aligned}
		\Delta \tau_n(t)\approx(n-1)^2\delta_{\mathrm{T}}^2\cos^2\psi_{m}/(2r_{m}c).
	\end{aligned}
\end{equation} 
To ensure~\eqref{DistanceSimp3Con}, we introduce a sample precision parameter $\epsilon_{\mathrm{t}}\ll 1$ and determine $t_{m}$ sequentially, i.e.,
\begin{equation}
	t_{m+1} \triangleq \min\{t_{m}+\epsilon_{\mathrm{t}}\Delta_{m},T\}.
\end{equation} 

According to \eqref{DistanceSimp1}, the Doppler shift can be expressed as $\tilde{\nu}_{m}\triangleq-(f_{\mathrm{c}}v/c)\sin(\alpha+\psi_{m})$, the simplified propagation delay can be expressed as $\tilde{\tau}_{m}\triangleq r_m/c$, and the array steering vector can be expressed as
\begin{equation}\label{ArraySteeringVector1}
	\begin{aligned}
		&\boldsymbol{a}(N_{\mathrm{T}},\psi_{m},f,r_{m})  \\
		&\!\triangleq\! \frac{1}{\sqrt{N_{\mathrm{T}}}}\!\Big[1,\ e^{-j\pi\frac{2\delta_{\mathrm{T}}}{\lambda_{\mathrm{c}}}\big(1+\frac{f}{f_{\mathrm{c}}}\big)\big(\sin\psi_{m}-\frac{\delta_{\mathrm{T}}}{2r_{m}}\cos^2\psi_{m}\big)},\!\dots\!,\\
		&\  \!e^{-j\pi\frac{2\delta_{\mathrm{T}}}{\lambda_{\mathrm{c}}}\big(1+\frac{f}{f_{\mathrm{c}}}\big)\big((N_{\mathrm{T}}-1)\sin\psi_{m}-\frac{\delta_{\mathrm{T}}}{2r_{m}}(N_{\mathrm{T}}-1)^2\cos^2\psi_{m}\big)}\!\Big]^\mathrm{T}\!.
	\end{aligned}
\end{equation}

Since the difference of the large-scale fading among different antennas is negligible, we have $\beta_{n}(t)\approx\beta(t)\triangleq\beta_{1}(t), n=2,3,\dots,N_{\mathrm{T}}$. Furthermore, as the duration of each sub-interval is small, we approximate $\beta(t)$ by $\beta(t)\approx\beta(t_m)=\sqrt{P_{\mathrm{T}}/P_{\mathrm{L}}(\psi_{m})}$ for $t\in[t_m,t_{m+1})$, where $P_{\mathrm{T}}$ and $P_{\mathrm{L}}(\psi_{m})$ represent the transmission power and path loss, respectively. In particular, $P_{\mathrm{L}}(\psi_{m})$ can be expressed as
\begin{equation}\label{Path_Loss}
	\begin{aligned}
		P_{\mathrm{L}}(\psi_{m})\!=\!\Big(\frac{4\pi r_{0}}{\lambda_{\mathrm{c}}}\Big)^2\Big(\frac{r_{m}}{r_{0}}\Big)^{\eta}\!=\!\frac{16\pi^2y_{0}^{\eta}\cos^{\eta}\alpha}{\lambda_{\mathrm{c}}^2r_0^{\eta-2}\cos^{\eta}(\psi_{m}+\alpha)},
	\end{aligned}
\end{equation}
where $r_0$ and $\eta$ represent the reference distance and path loss exponent, respectively. 

To further simplify the notation, we define
\begin{equation}\label{LargeScaleFading}
	\begin{aligned}
		\tilde{\beta}_{m}&\triangleq\sqrt{N_{\mathrm{T}}}\beta(t_m)e^{-j2\pi f_{\mathrm{c}}\tilde{\tau}_{m}}\\
		&=\sqrt{N_{\mathrm{T}}\tilde{P}_{\mathrm{T}}} e^{-j2\pi f_{\mathrm{c}}\tilde{\tau}_{m}}\cos^{\frac{\eta}{2}}(\psi_{m}+\alpha),
	\end{aligned}
\end{equation}
where $\tilde{P}_{\mathrm{T}}\triangleq P_{\mathrm{T}}\lambda_{\mathrm{c}}^2r_0^{\eta-2}/(16\pi^2y_{0}^{\eta}\cos^{\eta}\alpha)$. Then, the time-varying channel model  in the frequency domain can be established as
\begin{equation}\label{ChannelModel}
	\begin{aligned}
		\boldsymbol{h}(f,t)\!=\!\tilde{\beta}_{m} e^{-j2\pi f\tilde{\tau}_{m}} e^{j 2\pi \big(1+\frac{f}{f_{\mathrm{c}}}\big) \tilde{\nu}_{m}(t-t_{m})}\!\boldsymbol{a}^\mathrm{H}(N_{\mathrm{T}},\!\psi_{m},\!f,\!r_{m}).
	\end{aligned}
\end{equation}
Note that the distance-associated component is included in~\eqref{ChannelModel} because the minimum distance between the HST and BS can be as short as several meters~\cite{guanRealisticHighSpeedTrain2018}. As the HST approaches the BS, it may enter the near field of the ULA at the BS.

\subsection{Transition from Far-Field to Near-Field}
To determine the boundary between the far-field and near-field regions, we introduce the bandwidth-aware-near-field distance (BAND)~\cite{deshpandeWidebandGeneralization2023}. This metric is derived from the loss in the normalized beam gain due to the mismatch between the near-field wideband channel response and the optimal beamformer designed under far-field and narrowband assumptions, as given by
\begin{equation}\label{Lossfunction}
	\begin{aligned}
		\mathcal{L}(r,\psi,f,N_\mathrm{T}) &= 1 - \bigg\vert \frac{1}{N_\mathrm{T}} \sum_{n=1}^{N_\mathrm{T}}\!e^{-j2\pi \frac{f}{f_{\mathrm{c}}}(n-1)\frac{\delta_{\mathrm{T}}}{\lambda_{\mathrm{c}}}\sin\psi}\\
		&\quad e^{j2\pi\big((1+\frac{f}{f_{\mathrm{c}}})\frac{(n-1)^2\delta_{\mathrm{T}}^2\cos^2\psi}{2 r \lambda_{\mathrm{c}}}\big)} \bigg\vert,
	\end{aligned}
\end{equation}
where $r>0$ is a variable representing the distance. Given $\psi$, $N_\mathrm{T}$, and $B_\mathrm{f}$, the BAND is defined as the smallest distance beyond which the loss $\mathcal{L}(r,\psi,f,N_\mathrm{T})$ is no larger than a given threshold $\mathcal{L}_{\mathrm{th}}$ for any $f\in [-B_\mathrm{f}/2,B_\mathrm{f}/2]$. Specifically, during $[t_{m},t_{m+1})$, the BAND, denoted as $R_{\mathrm{B}}(\psi_{m},N_\mathrm{T},B_{\mathrm{f}})$, can be obtained by solving the optimization problem
\begin{equation}\label{BAND}
		\min_{r>0}\ r,\quad
		\mathrm{s.t.}\ \mathcal{L}(r,\psi_{m},f,N_\mathrm{T})\!\leq\! \mathcal{L}_{\mathrm{th}},\ \forall f\in \!\Big[\!-\!\frac{B_{\mathrm{f}}}{2},\frac{B_{\mathrm{f}}}{2}\!\Big].
\end{equation}

Due to page limits, we only consider the beam design for the narrowband scenario in this paper, leaving the wideband scenario for future work. Given that $B_{\mathrm{f}}\ll f_{\mathrm{c}}$ and $f/f_{\mathrm{c}}\approx 0,\ \forall f\in[-B_\mathrm{f}/2,B_\mathrm{f}/2]$, the BAND $R_{\mathrm{B}}(\psi_{m},N_\mathrm{T},B_\mathrm{f})$ can be approximated by $R_{\mathrm{B}}(\psi_{m},N_\mathrm{T},0)$. Thus, if $r_{m}>R_{\mathrm{B}}(\psi_{m},N_\mathrm{T},0)$, the HST is considered to be in the far field of the ULA at the BS; otherwise, the HST is considered to be in the near field. Note that although the boundary between the far-field and near-field regions is determined by the BAND, no further approximations are made for the channel samples in the far-field regions. All channel samples used in the following sections are generated according to the near-field channel model in~\eqref{ChannelModel}.

\subsection{RSNR}
Let $\hat{\psi}_{1},\hat{\psi}_{2},\dots,\hat{\psi}_{M}$ denote a series of estimated AoDs corresponding to the true AoDs $\psi_{1},\psi_{2},\dots,\psi_{M}$. During the downlink transmission, the BS switches beams according to the estimated HST location $\boldsymbol{r}(\hat{\psi}_{m}),\ m\in\mathcal{I}_{\mathrm{M}}$. If $\hat{\psi}_{m}\in[\varphi_{i}, \varphi_{i+1})$, the BS transmits the $i$th beam during $[t_{m},t_{m+1})$, and the normalized beam gain can be defined as
\begin{equation}\label{BeamformingGain}
	\begin{aligned}
		g_{m}(\boldsymbol{f}_{i})\triangleq\big\vert\boldsymbol{a}^\mathrm{H}\big(N_{\mathrm{T}},\psi_{m},0,r_{m}\big)\boldsymbol{f}_{i}\big\vert^2 =\boldsymbol{f}_{i}^{\mathrm{H}}\boldsymbol{A}_{m}\boldsymbol{f}_{i},
	\end{aligned}
\end{equation}
where $\boldsymbol{A}_{m}\triangleq\boldsymbol{a}\big(N_{\mathrm{T}},\psi_{m},0,r_{m}\big)\boldsymbol{a}^\mathrm{H}\big(N_{\mathrm{T}},\psi_{m},0,r_{m}\big)$. The corresponding RSNR can be expressed as
\begin{equation}\label{ReceivingSNR}
	\Gamma_{m}(\boldsymbol{f}_{i})=\frac{\big\vert\tilde{\beta}_{m}\big\vert^2}{P_{\mathrm{N}}}\boldsymbol{f}_{i}^{\mathrm{H}}\boldsymbol{A}_{m}\boldsymbol{f}_{i},
\end{equation}
where $P_{\mathrm{N}}$ represents the noise power. Considering all possible relationships between $\hat{\psi}_{m}$ and $\varphi_{i},\ i=1,2,\dots,N+1$, we can express the instantaneous RSNR as
\begin{equation}\label{QoS2}
	\widetilde{\Gamma}_{m}\big(\{\boldsymbol{f}_{i}\}_{i=1}^{N}\big)=\sum_{i=1}^{N}\Gamma_{m}(\boldsymbol{f}_{i})\big(\varepsilon(\hat{\psi}_{m}-\varphi_{i})-\varepsilon(\hat{\psi}_{m}-\varphi_{i+1})\big),
\end{equation}
where $\varepsilon(\psi)$ is a step function.

\vspace{0.5mm}
\section{Sequential Beam Design}\label{ProblemFormulation}
In this paper, we consider beam switching based beam design, aiming to minimize $N$ within a predetermined railway range from $\boldsymbol{r}(\psi_{\mathrm{min}})$ to $\boldsymbol{r}(\psi_{\mathrm{max}})$, subject to constant modulus constraints and RSNR constraints that require the RSNR at each HST location no lower than a predetermined threshold $\gamma_{\mathrm{th}}$. Specifically, according to~\eqref{QoS2}, there are two possibilities for each beam switching: if $\varphi_{i} \leq \psi_{m} <\varphi_{i+1}$, the HST is within the angle coverage of $\boldsymbol{f}_{i}$ as scheduled; otherwise, the estimation error in AoD leads to beam misalignment. To avoid the degradation of RSNR caused by the misalignment, we impose constraints on the RSNR for AoD samples estimated to fall into $[\varphi_{i}, \varphi_{i+1})$ with a probability greater than a threshold $p_{\mathrm{th}}$. For simplicity, we assume that the estimation error in each AoD sample independently follows a Gaussian distribution. Given $\psi_{m}$, the estimated AoD can be expressed as $\hat{\psi}_{m}=\psi_{m}+\varepsilon_{\psi,m}$, where $\varepsilon_{\psi,m}\sim\mathcal{N}(0,\sigma_\psi^2)$, and the probability that $\hat{\psi}_{m}$ falls into $[\varphi_{i}, \varphi_{i+1})$ can be expressed as
\begin{equation}\label{Probability}
	\begin{aligned}
		P\big\{\varphi_{i}\!\le\!\hat{\psi}_{m}\!<\! \varphi_{i+1}\big\vert\psi_{m}\big\}\!=\!Q\Big(\!\frac{\varphi_{i}-\psi_{m}}{\sigma_{\psi}}\!\Big)-Q\Big(\!\frac{\varphi_{i+1}\!-\!\psi_{m}}{\sigma_{\psi}}\!\Big).
	\end{aligned}
\end{equation}
The set of AoD samples with $P\{\varphi_{i}\le\hat{\psi}_{m}< \varphi_{i+1}\vert\psi_{m}\}\geq p_{\mathrm{th}}$ can be expressed as
\begin{equation}\label{Index_set}
	\mathcal{M}(\varphi_{i},\varphi_{i+1})=\big\{m\in \mathcal{I}_{\mathrm{M}}\big\vert P\{\varphi_{i}\le\hat{\psi}_{m}< \varphi_{i+1}\vert\psi_{m}\}\geq p_{\mathrm{th}}\big\}.
\end{equation}
For any $m$ in $\mathcal{M}(\varphi_{i},\varphi_{i+1})$, we require the RSNR no lower than $\gamma_{\mathrm{th}}$, i.e., 
\begin{equation}\label{QoS1}
	\Gamma_{m}(\boldsymbol{f}_{i})\ge \gamma_{\mathrm{th}},\ \forall m\in\mathcal{M}(\varphi_{i},\varphi_{i+1}). 
\end{equation}
Note that with a sufficiently low $p_{\mathrm{th}}$, the RSNR constraint in~\eqref{QoS1} ensures overlapping angular regions between neighboring beams. If the estimation for $\psi_{m}$ is accurate, i.e., $\hat{\psi}_{m}\approx \psi_{m}$, the AoD sample set $\mathcal{M}(\varphi_{i},\varphi_{i+1})$ can be simplified into
\begin{equation}\label{Index_set1}
	\mathcal{M}(\varphi_{i},\varphi_{i+1})=\big\{m\in \mathcal{I}_{\mathrm{M}}\big\vert \varphi_{i} \leq \psi_{m} <\varphi_{i+1} \big\}.
\end{equation}
For simplicity, we assume $\hat{\psi}_{m}\approx \psi_{m}$ and use~\eqref{Index_set1} in the following sections. Nonetheless, the beam design schemes proposed in this paper can be extended to cases where the estimation error is not negligible by replacing \eqref{Index_set1} with \eqref{Index_set}.

Under the assumption of accurate AoD estimation, the optimization problem aiming at minimizing $N$, subject to~\eqref{Constant_Constraint} and~\eqref{QoS1}, can be expressed as
\begin{equation}\label{Optimization_Origin}
	\begin{aligned}
		\min_{N,\{\boldsymbol{f}_{i}\}_{i=1}^{N},\{\varphi_{i}\}_{i=2}^{N+1}}&N\\
		\mathrm{s.t.}\quad\quad\ \ &\eqref{Constant_Constraint},~\eqref{QoS1},\ \forall i\!=\!1,2,\dots,N\\
		&\!\varphi_{i+1}\!>\!\varphi_{i},\ \forall i\!=\!1,2,\dots,N\\
		&\!\varphi_{i+1}\!\in\!\{\psi_{m}\}_{m\in \mathcal{I}_{\mathrm{M}}},\ \forall i\!=\!1,2,\dots,N\\
		&\!\varphi_{N+1}\geq\psi_{\mathrm{max}}.
	\end{aligned}
\end{equation}
Note that~\eqref{Optimization_Origin} is difficult to solve due to the large number of non-convex constraints and the interdependence between $N$, $\{\boldsymbol{f}_{i}\}_{i=1}^{N}$, and $\{\varphi_{i}\}_{i=2}^{N+1}$. Therefore, we propose a sequential beam design approach to convert the problem into a series of beam design sub-problems that aim at maximizing
the beam coverage subject to~\eqref{Constant_Constraint} and~\eqref{QoS1}, where the beam coverage is denoted as $\varphi_{i+1}-\varphi_{i}$. According to~\eqref{ReceivingSNR}, $\gamma_{\mathrm{th}}$ can be equivalently transformed into the following beam gain threshold
\begin{equation}\label{Gamma}
	\gamma_{m}\triangleq\frac{\gamma_{\mathrm{th}}\sigma_{\mathrm{N}}^2}{\big\vert\tilde{\beta}_{m}\big\vert^2}=\frac{\gamma_{\mathrm{th}}P_{\mathrm{N}}}{N_{\mathrm{T}}\tilde{P}_{\mathrm{T}}\cos^{\eta}(\psi_{m}+\alpha)},\ \forall m\in\mathcal{I}_{\mathrm{M}}.
\end{equation}
For any $\psi_{m}\in[\varphi_{i},\varphi_{i+1})$,~\eqref{QoS1} is equivalent to 
\begin{equation}\label{QoS}
	\boldsymbol{f}_{i}^{\mathrm{H}}\boldsymbol{A}_{m}\boldsymbol{f}_{i}\geq \gamma_{m},\ \forall m\in\mathcal{M}(\varphi_{i},\varphi_{i+1}).
\end{equation}
Then, for the $i$th beam, with $\varphi_i$ fixed, the sub-problem can be expressed as
\begin{equation}\label{Optimization}
	\begin{aligned}
		\max_{\boldsymbol{f}_{i},\varphi_{i+1}}&\quad\varphi_{i+1}\\
		\mathrm{s.t.}&\quad\eqref{Constant_Constraint},~\eqref{QoS},\ \varphi_{i+1}>\varphi_{i},\ \varphi_{i+1}\in\{\psi_{m}\}_{m\in \mathcal{I}_{\mathrm{M}}}.
	\end{aligned}
\end{equation}
The design of the next beam can be performed through updating $i\leftarrow i+1$ and solving~\eqref{Optimization} again. The proposed sequential beam design approach is summarized in~\textbf{Algorithm~\ref{alg1}}, with $\big\{\hat{\boldsymbol{f}}_i\big\}_{i=1}^{N}$ and $\big\{\hat{\varphi}_i\big\}_{i=2}^{N+1}$ representing the results from sequentially solving~\eqref{Optimization}. Once $\hat{\varphi}_{i}\geq \psi_{\mathrm{max}}$, the sequential beam design stops.

Note that~\eqref{Constant_Constraint} and~\eqref{QoS} lead~\eqref{Optimization} to be non-convex and NP-hard. In particular,~\eqref{QoS} can be reformulated as $\gamma_{m}\big(\varepsilon(\psi_{m}-\varphi_{i})-\varepsilon(\psi_{m}-\varphi_{i+1})\big)-\boldsymbol{f}_{i}^\mathrm{H}\boldsymbol{A}_{m}\boldsymbol{f}_{i}\leq0,\ \forall m\in \mathcal{I}_{\mathrm{M}}$, where $\varepsilon(\psi)$ is non-convex and incontinuous, posing challenges for the application of existing optimization methods. Therefore, we consider a TS framework, where $\varphi_{i+1}$ and $\boldsymbol{f}_{i}$ are optimized alternately. In the first stage, by fixing $\varphi_{i+1}$, we transform~\eqref{Optimization} into a feasibility problem to optimize $\boldsymbol{f}_{i}$. In the second stage, we determine $\varphi_{i+1}$ according to the output of the first stage using search methods. The two stages are performed alternately until the largest achievable $\varphi_{i+1}$ is found. Specifically, we denote the optimal solution of $\varphi_{i+1}$ in~\eqref{Optimization} as $\varphi_{i+1}^{\star}$ and fix $\varphi_{i+1}$ as $\varphi_{i+1}^{(p)}$ in the $p$th TS iteration, where $p$ is a non-negative integer. With $\varphi_{i+1}$ fixed,~\eqref{Optimization} can be transformed into a feasibility problem as
\begin{equation}\label{Feasibility}
	\begin{aligned}
		\mathrm{find}&\quad\boldsymbol{f}_{i} \\
		\mathrm{s.t.}&\quad \eqref{Constant_Constraint},\ \boldsymbol{f}_{i}^{\mathrm{H}}\boldsymbol{A}_{m}\boldsymbol{f}_{i}\geq \gamma_{m},\  m\in\mathcal{M}(\varphi_{i},\varphi_{i+1}^{(p)}).
	\end{aligned}
\end{equation}
If a feasible solution can be found from~\eqref{Feasibility}, we have $\varphi_{i+1}\leq\varphi_{i+1}^{\star}$ so that $\varphi_{i+1}^{(p+1)}>\varphi_{i+1}^{(p)}$ is set to adjust $\varphi_{i+1}$ upwards in the next iteration; otherwise, we have $\varphi_{i+1}>\varphi_{i+1}^{\star}$, and set $\varphi_{i+1}^{(p+1)}<\varphi_{i+1}^{(p)}$. Once the sequence $\big\{\varphi_{i+1}^{(p)}\big\}$ converges, the TS iterations stop.

\begin{algorithm}[!t]
	\caption{Sequential Beam Design}\label{alg:alg1}
	\begin{algorithmic}[1]
		\STATE \emph{Input:}  $\psi_{\mathrm{min}}$, $\psi_{\mathrm{max}}$, $N_{\mathrm{T}}$, $\epsilon_{\mathrm{max}}$, $\epsilon_{\mathrm{min}}$, $\epsilon_{\mathrm{u}}$, $\epsilon_{\psi}$, $\epsilon_{1}$, $\Delta \rho_{2}$, $\Delta\varphi$, $w$, $\{\gamma_m\}_{m\in \mathcal{I}_{\mathrm{M}}}$, $\{\boldsymbol{A}_{m}\}_{m\in \mathcal{I}_{\mathrm{M}}}$, $\{\psi_m\}_{m\in \mathcal{I}_{\mathrm{M}}}$.
		\STATE Set $i \leftarrow 1$, $\hat{\varphi}_{1}\leftarrow\psi_{\mathrm{min}}$ and $\varphi_{1}\leftarrow\hat{\varphi}_{1}$.
		\WHILE {$\hat{\varphi}_{i} < \psi_{\mathrm{max}}$}
		\STATE Obtain $\hat{\boldsymbol{f}}_{i}$ and $\hat{\varphi}_{i+1}$ through \textbf{Algorithms~\ref{alg2}} or~\textbf{\ref{alg4}}.
		\STATE Update $i\leftarrow i+1$ and set $\varphi_{i}\leftarrow\hat{\varphi}_{i}$
		\ENDWHILE 
		\STATE Obtain $N\leftarrow i-1$.
		\STATE \emph{Output:} $N$, $\big\{\hat{\boldsymbol{f}}_i\big\}_{i=1}^{N}$, $\big\{\hat{\varphi}_i\big\}_{i=2}^{N+1}$.
	\end{algorithmic}
	\label{alg1}
\end{algorithm}

To solve~\eqref{Optimization}, we propose two beam design schemes under the TS framework. The first scheme is named SDR-DC-BiS and is summarized in~\textbf{Algorithm~\ref{alg2}}, where the problem in the first stage is solved based on SDR and a DC algorithm, and in the second stage the beam coverage is determined by a BiS method. The second scheme is named PP-PDG-MS and is summarized in~\textbf{Algorithm~\ref{alg4}}, where the problem in the first stage is solved based on a PP method and a PDG algorithm with closed-form solutions, and in the second stage the beam coverage is determined by a MS method. In fact, the SDR-DC-BiS scheme is suitable for the cases that $N_{\mathrm{T}}$ is small and the number of constraints is not large, while the PP-PDG-MS scheme is suitable for the cases that either $N_{\mathrm{T}}$ or the number of constraints is large. 

\section{SDR-DC-BiS Scheme}\label{Method}
Under the TS framework, we propose the SDR-DC-BiS scheme. In the first stage, through transforming $\boldsymbol{f}_{i}$ into a semidefinite positive Hermitian matrix and introducing an additional rank-one constraint, we establish an equivalent reformulation of~\eqref{Feasibility}. To address the non-convexity brought by the rank-one constraint and the potential infeasibility brought by the RSNR constraints, we further transform the problem into a min-max and DC programming problem, which is then tackled by solving a series of convex sub-problems. In the second stage, $\varphi_{i+1}$ is determined using the BiS method. These two stages are alternately performed until a stop condition is triggered. Moreover, we also provide some analysis on the computational complexity of the SDR-DC-BiS scheme.

\subsection{SDR and DC Algorithm}
To tackle the non-convexity of the constraints in~\eqref{Feasibility}, based on the SDR method, we define $\boldsymbol{F}_{i}\in\mathbb{C}^{N_{\mathrm{T}}\times N_{\mathrm{T}}}$ with $\boldsymbol{F}_{i}\succeq 0$ and $\mathrm{rank}(\boldsymbol{F}_{i})=1$, and then transform~\eqref{Feasibility} into
\begin{subequations}\label{Feasibility_SDR}
\begin{align}
	\mathrm{find}&\quad\boldsymbol{F}_{i} \\
	\mathrm{s.t.}&\quad \mathrm{Tr}\{\boldsymbol{A}_{m}\boldsymbol{F}_{i}\}\geq\gamma_{m},\ \forall m\in\mathcal{M}(\varphi_{i},\varphi_{i+1}^{(p)})\label{Feasibility_SDR_QoS}\\
	&\quad [\boldsymbol{F}_{i}]_{n,n}= \frac{1}{N_{\mathrm{T}}}, \forall n\in\mathcal{I}_{\mathrm{N}}\label{Feasibility_SDR_ConMod}\\
	&\quad \boldsymbol{F}_{i}\succeq 0\label{Feasibility_SDR_PSD}\\
	&\quad \mathrm{rank}(\boldsymbol{F}_{i})=1.\label{Feasibility_SDR_Rank1}
\end{align}
\end{subequations}
Note that~\eqref{Feasibility_SDR_Rank1} ensures the equivalence between~\eqref{Feasibility} and~\eqref{Feasibility_SDR} but introduces the non-convexity. One method to address this issue is to remove~\eqref{Feasibility_SDR_Rank1} and relax~\eqref{Feasibility_SDR} into a convex problem. The relaxation simplifies the problem but breaks the equivalence. When the solution of the relaxed problem is not rank-one, additional processing procedures, such as randomization, are required to regenerate a feasible solution for~\eqref{Feasibility}, leading to high computational complexity. Moreover, the feasibility of~\eqref{Feasibility} can not be guaranteed even when the relaxed problem is feasible. All the above-mentioned problems motivate us to explore better methods.

Another method to address~\eqref{Feasibility_SDR_Rank1} is transforming it using a DC function~\cite{fuReconfigurableIntelligent2021c,maPassiveBeamforming2022}. Specifically, for any positive semidefinite matrix $\boldsymbol{F}_{i}$ with $\mathrm{Tr}\{\boldsymbol{F}_{i}\}>0$, we have the following equivalence
\begin{equation}
	\mathrm{rank}(\boldsymbol{F}_{i})=1 \Longleftrightarrow \Vert\boldsymbol{F}_{i}\Vert_{*}-\Vert\boldsymbol{F}_{i}\Vert_{2}=0,
\end{equation}
where $\Vert\boldsymbol{F}_{i}\Vert_{*}$ represents the sum of all singular values of $\boldsymbol{F}_{i}$ and $\Vert\boldsymbol{F}_{i}\Vert_{2}$ represents the maximum singular value of $\boldsymbol{F}_{i}$. Note that we also have $\mathrm{Tr}\{\boldsymbol{F}_{i}\}=\Vert\boldsymbol{F}_{i}\Vert_{*}$ for any $\boldsymbol{F}_{i}$ that satisfies~\eqref{Feasibility_SDR_PSD}, and $\mathrm{Tr}\{\boldsymbol{F}_{i}\}=1$ for any $\boldsymbol{F}_{i}$ that satisfies~\eqref{Feasibility_SDR_ConMod}. Therefore, with~\eqref{Feasibility_SDR_ConMod} and~\eqref{Feasibility_SDR_PSD},~\eqref{Feasibility_SDR_Rank1} is equivalent to $1-\Vert\boldsymbol{F}_{i}\Vert_{2}=0$. Then, we can reformulate~\eqref{Feasibility_SDR} as
\begin{equation}\label{Feasibility_SDR_Rank}
\begin{aligned}
	\min_{\boldsymbol{F}_{i}\in \mathcal{F}_{1}}\ -\rho_{1}\Vert\boldsymbol{F}_{i}\Vert_{2} \quad\mathrm{s.t.}\ \eqref{Feasibility_SDR_QoS},
\end{aligned}
\end{equation}
where $\rho_{1}$ is a penalty parameter, and
\begin{equation}
	\mathcal{F}_{1}\!\triangleq\!\big\{\boldsymbol{F}\!\in\! \mathbb{C}^{N_{\mathrm{T}}\times N_{\mathrm{T}}}\big\vert \boldsymbol{F}\!\succeq\!0,[\boldsymbol{F}]_{n,n}\!=\!1/N_{\mathrm{T}}, \forall n\in\mathcal{I}_{\mathrm{N}}\!\big\}.
\end{equation}
Note that~\eqref{Feasibility_SDR_Rank} is a penalty problem of~\eqref{Feasibility_SDR}. If the feasible set of~\eqref{Feasibility_SDR} is not empty and $\rho_{1}>0$, the optimal solution set of~\eqref{Feasibility_SDR_Rank} equals the feasible set of~\eqref{Feasibility_SDR}~\cite{lethiExactpenalty2012}.

Since the transmission power is limited, the feasible set of~\eqref{Feasibility_SDR_Rank} may be empty if $\varphi_{i+1}$ is set too large, posing challenges for the application of existing optimization methods. Therefore, we rewrite~\eqref{Feasibility_SDR_QoS} and transform~\eqref{Feasibility_SDR_Rank} into an optimization problem that is always feasible. To facilitate the transformation, we consider the equivalence between the following two inequalities
\begin{equation}\label{Relax_QoD}
	\eqref{Feasibility_SDR_QoS}\!\Longleftrightarrow\!-\frac{1}{\gamma_{m}}\mathrm{Tr}\{\boldsymbol{A}_{m}\boldsymbol{F}_{i}\}\!\leq\! -1,\ \!\forall m\in\!\mathcal{M}(\varphi_{i},\varphi_{i+1}^{(p)}).
\end{equation}
Based on the equivalence, we transform~\eqref{Feasibility_SDR_Rank} into
\begin{equation}\label{Feasibility_SDR_Rank_Penalty}
	\min_{\boldsymbol{F}_{i}\in \mathcal{F}_{1}}\max_{m\in\mathcal{M}(\varphi_{i},\varphi_{i+1}^{(p)})}-\frac{1}{\gamma_{m}}\mathrm{Tr}\big\{ \boldsymbol{A}_{m}\boldsymbol{F}_{i}\big\}-\rho_{1}\Vert\boldsymbol{F}_{i}\Vert_{2}.
\end{equation}
To simplify the notation, we define
\begin{equation}
	D_{1}(\boldsymbol{F}_{i})\triangleq\max_{m\in\mathcal{M}(\varphi_{i},\varphi_{i+1}^{(p)})}-\frac{1}{\gamma_{m}}\mathrm{Tr}\big\{ \boldsymbol{A}_{m}\boldsymbol{F}_{i}\big\},\ \boldsymbol{F}_{i}\succeq 0,
\end{equation}
and $D_{2}(\boldsymbol{F}_{i})\triangleq\rho_{1}\Vert\boldsymbol{F}_{i}\Vert_{2},\ \boldsymbol{F}_{i}\succeq 0.$
The objective function in~\eqref{Feasibility_SDR_Rank_Penalty} can then be rewritten as $D(\boldsymbol{F}_{i})\triangleq D_{1}(\boldsymbol{F}_{i})-D_{2}(\boldsymbol{F}_{i}),\ \boldsymbol{F}_{i}\succeq 0$. Note that~\eqref{Feasibility_SDR_Rank_Penalty} is the penalty problem of
\begin{equation}\label{Feasibility_SDR_QoS_Penalty}
	\min_{\boldsymbol{F}_{i}\in\mathcal{F}_{1}}\ D_{1}(\boldsymbol{F}_{i})\quad \mathrm{s.t.}\ 1-\Vert\boldsymbol{F}_{i}\Vert_{2}=0.
\end{equation}
If $\rho_{1}$ is sufficiently large,~\eqref{Feasibility_SDR_Rank_Penalty} and~\eqref{Feasibility_SDR_QoS_Penalty} have the same optimal solution sets~\cite{lethiExactpenalty2012}. Moreover, if $D_{2}(\boldsymbol{F}_{i}^{\star})=\rho_{1}$ and $D_{1}(\boldsymbol{F}_{i}^{\star})\leq -1$, where $\boldsymbol{F}_{i}^{\star}$ is an optimal solution of~\eqref{Feasibility_SDR_Rank_Penalty}, then $\boldsymbol{F}_{i}^{\star}$ is a feasible solution of~\eqref{Feasibility_SDR}; otherwise, the feasible set of~\eqref{Feasibility_SDR} is empty.

Since both $D_{1}(\boldsymbol{F}_{i})$ and $D_{2}(\boldsymbol{F}_{i})$ are convex functions,~\eqref{Feasibility_SDR_Rank_Penalty} is a typical case of a DC programming problem, which allows the application of existing DC algorithms. The DC algorithm involves the construction of two coupled sequences $\big\{ \boldsymbol{F}_{i}^{(q)}\big\}$ and $\big\{\boldsymbol{\Xi}_{i}^{(q)}\big\}$~\cite{wenProximaldifferenceofconvex2018}, where $q$ is the index of DC iteration. For the $q$th iteration, $\boldsymbol{F}_{i}^{(q)}$ is a feasible solution of~\eqref{Feasibility_SDR_Rank_Penalty}, while $\boldsymbol{\Xi}_{i}^{(q)}$ is an element chosen from the subdifferential of $D_{2}(\boldsymbol{F}_{i})$ at $\boldsymbol{F}_{i}^{(q)}$, i.e., $\boldsymbol{\Xi}_{i}^{(q)}\in\partial D_{2}(\boldsymbol{F}_{i}^{(q)})$. Since $D_{2}(\boldsymbol{F}_{i})$ is convex and continuously differentiable, $\partial D_{2}(\boldsymbol{F}_{i}^{(q)})$ reduces to the gradient of $D_{2}(\boldsymbol{F}_{i})$ at $\boldsymbol{F}_{i}^{(q)}$, and $\boldsymbol{\Xi}_{i}^{(q)}$ can be given by 
\begin{equation}\label{Subdifferential_Rank_Matrix}
	\boldsymbol{\Xi}_{i}^{(q)} = \rho_{1}\boldsymbol{\xi}_{i}^{(q)}\big(\boldsymbol{\xi}_{i}^{(q)}\big)^{\mathrm{H}},
\end{equation}
where $\boldsymbol{\xi}_{i}^{(q)}$ is the eigenvector corresponding to the largest eigenvalue of $\boldsymbol{F}_{i}^{(q)}$. Based on~\eqref{Subdifferential_Rank_Matrix}, we establish a surrogate function for $D(\boldsymbol{F}_{i})$ as
\begin{equation}
	\hat{D}\big(\!\boldsymbol{F}_{i},\boldsymbol{F}_{i}^{(q)} \big)\!\!\triangleq\!D_{1}(\boldsymbol{F}_{i})-D_{1}(\boldsymbol{F}_{i}^{(q)}) -\!\mathrm{Tr}\big\{\!\boldsymbol{\Xi}_{i}^{(q)}\!(\boldsymbol{F}_{i}-\!\boldsymbol{F}_{i}^{(q)})\!\big\}\!.
\end{equation}
Then, $\boldsymbol{F}_{i}^{(q+1)}$ can be obtained through solving the following convex sub-problem
\begin{equation}\label{DC_Subproblem}
	\boldsymbol{F}_{i}^{(q+1)}= \arg\min_{\boldsymbol{F}_{i}\in\mathcal{F}_{1}}\ \hat{D}\big(\boldsymbol{F}_{i},\boldsymbol{F}_{i}^{(q)}\big).
\end{equation}

Note that according to the definition of $\boldsymbol{\Xi}_{i}^{(q)}$, we have
\begin{equation}
	\begin{aligned}
		D(\boldsymbol{F}_{i}^{(q+1)})&\leq  \hat{D}\big(\boldsymbol{F}_{i}^{(q+1)},\boldsymbol{F}_{i}^{(q)}\big)=\min_{\boldsymbol{F}_{i}\in \mathcal{F}_{1}}\hat{D}\big(\boldsymbol{F}_{i},\boldsymbol{F}_{i}^{(q)}\big) \\
		&\leq \hat{D}\big(\boldsymbol{F}_{i}^{(q)},\boldsymbol{F}_{i}^{(q)}\big)=D(\boldsymbol{F}_{i}^{(q)}),
	\end{aligned}
\end{equation}
which means that $\big\{ D(\boldsymbol{F}_{i}^{(q)})\big\}$ is non-increasing. Furthermore, since $D(\boldsymbol{F}_{i})$ is lower bounded, $\big\{ U(\boldsymbol{F}_{i}^{(q)})\big\}$ can generally converge to a local minimum~\cite{sunMajorizationMinimizationAlgorithms2017}. 

Although the problem of~\eqref{DC_Subproblem} is convex, it is still difficult to solve because $D_{1}(\boldsymbol{F}_{i})$ is non-smooth. One method to address this issue is to introduce an auxiliary variable, denoted as $\mu_{\mathrm{f}}$, and transform the problem of~\eqref{DC_Subproblem} into
\begin{subequations}\label{DC_Subproblem_Smooth}
	\begin{align}
		\min_{\boldsymbol{F}_{i}\in \mathcal{F}_{1},\mu_{\mathrm{f}}}&\ \mu_{\mathrm{f}}-\mathrm{Tr}\big\{\boldsymbol{\Xi}_{i}^{(q)}\boldsymbol{F}_{i}\big\} \\
		\mathrm{s.t.}\ \ &\ \!-\!\frac{1}{\gamma_{m}}\mathrm{Tr}\big\{ \!\boldsymbol{A}_{m}\boldsymbol{F}_{i}\big\}\!\leq\! \mu_{\mathrm{f}},\ \forall m\in\!\mathcal{M}(\varphi_{i},\!\varphi_{i+1}^{(p)}).\label{DC_Subproblem_Smooth_QoS}
	\end{align}
\end{subequations}
Note that~\eqref{DC_Subproblem_Smooth} is equivalent to the problem in~\eqref{DC_Subproblem} and can be easily solved by CVX. However, this smoothing method introduces a large number of constraints, resulting in high computational complexity, as will be discussed in Section~\ref{Complexity}. 

The SDR and DC algorithm is presented from step 9 to step 14 in \textbf{Algorithm~\ref{alg2}}. The iterations stop once the following conditions are satisfied, i.e.,
\begin{equation}\label{DC_Stop_Conditions}
	D(\boldsymbol{F}_{i}^{(q-1)})-D(\boldsymbol{F}_{i}^{(q)}) \leq\epsilon_{1},1-\big\Vert\boldsymbol{F}_{i}^{(q)}\big\Vert_2 \leq\epsilon_{2},
\end{equation}
where $\epsilon_{1}$ and $\epsilon_{2}$ are predetermined error tolerance thresholds. The first condition in~\eqref{DC_Stop_Conditions} is defined to ensure the convergence of $D(\boldsymbol{F}_{i})$, whereas the second is defined to limit the degree of violation to~\eqref{Feasibility_SDR_Rank1}. 

\subsection{Bisection Search}\label{Bisection_Search}
In the second stage, we determine $\varphi_{i+1}$ through BiS based on the result from the first stage. Suppose $\big\{ \boldsymbol{F}_{i}^{(q)}\big\}$ converges at $\boldsymbol{F}_{i}^{(q_{\star})}$. Ideally, if $D_1(\boldsymbol{F}_{i}^{(q_{\star})})\leq -1$ and $D_2(\boldsymbol{F}_{i}^{(q_{\star})})=\rho_{1}$, $\boldsymbol{F}_{i}^{(q_{\star})}$ is a feasible solution of~\eqref{Feasibility_SDR}, and therefore $\varphi_{i+1}$ should be increased; otherwise, $\varphi_{i+1}$ should be decreased. 

Note that the computational complexity of the first stage may become intolerable if $\varphi_{i+1}$ is set too large. Moreover, only the solution from the final TS iteration is useful, making the high complexity unnecessary. To enhance the efficiency, we generate a lower bound for $D_{1}(\boldsymbol{F}_{i})$ after
each adjustment of $\varphi_{i+1}$ through solving the following SDR problem
\begin{equation}\label{DC_SDR_Initialization}
	\begin{aligned}
		\min_{\boldsymbol{F}_{i}\in \mathcal{F}_{1},\mu_{\mathrm{f}}}\ \mu_{\mathrm{f}}\quad\mathrm{s.t.}\ \eqref{DC_Subproblem_Smooth_QoS}.
	\end{aligned}
\end{equation}
Denote the solution of~\eqref{DC_SDR_Initialization} as $\boldsymbol{F}_i^{(0)}$. Since the rank-one constraint is removed from~\eqref{DC_SDR_Initialization}, $D_1(\boldsymbol{F}_{i}^{(0)})$ provides a lower bound for $D_1(\boldsymbol{F}_{i}^{(q_{\star})})$. If $D_1(\boldsymbol{F}_{i}^{(0)})>-1$, we immediately have $D_1(\boldsymbol{F}_{i}^{(q_{\star})})>-1$, and therefore the computational complexity of the first stage can be reduced. If $D_1(\boldsymbol{F}_{i}^{(0)})\leq-1$, $\boldsymbol{F}_{i}^{(0)}$ provides a good initialization for the first stage through setting $\boldsymbol{\Xi}_{i}^{(0)} = \rho_{1}\boldsymbol{\xi}_{i}^{(0)}\big(\boldsymbol{\xi}_{i}^{(0)}\big)^{\mathrm{H}}$, and the number of DC iterations can be reduced. 

In numerical simulations, it is nearly impossible to achieve the exact equality of $D_2(\boldsymbol{F}_{i}^{(q_{\star})})= \rho_{1}$. To determine the feasibility in a more practical way, we establish a beamformer based on $\boldsymbol{F}_{i}^{(q_{\star})}$ as
\begin{equation}\label{Optimization_solution1}
	\big[\boldsymbol{f}_i^{(q_{\star})}\big]_{n} = \frac{1}{\sqrt{N_{\mathrm{T}}}}\frac{[\boldsymbol{\xi}_{i}^{(q_{\star})}]_{n}}{\big\vert[\boldsymbol{\xi}_{i}^{(q_{\star})}]_{n}\big\vert},\ \forall n\in\mathcal{I}_{\mathrm{N}}.
\end{equation}
If $\boldsymbol{f}_i^{(q_{\star})}$ is feasible for~\eqref{Feasibility}, we set $\varphi_{i+1}^{\mathrm{lb}}=\varphi_{i+1}^{(p)}$ and $\hat{\boldsymbol{f}}_{i}=\boldsymbol{f}_i^{(q_{\star})}$; otherwise, we set $\varphi_{i+1}^{\mathrm{ub}}=\varphi_{i+1}^{(p)}$, where $[\varphi_{i+1}^{\mathrm{lb}},\varphi_{i+1}^{\mathrm{ub}}]$ represents the search interval for $\varphi_{i+1}$. Then, we increment $p$ by 1 and update the AoD for the next TS iteration as $\varphi_{i+1}^{(p)} \leftarrow (\varphi_{i+1}^{\mathrm{lb}} + \varphi_{i+1}^{\mathrm{ub}})/2$. BiS stops the iterations once $\varphi_{i+1}^{\mathrm{ub}}-\varphi_{i+1}^{\mathrm{lb}}\leq\epsilon_{\varphi}$, where $\epsilon_{\varphi}$ is a predetermined threshold. Suppose BiS stops at $p=p_{\star}$. Then $\hat{\varphi}_{i+1}$ can be expressed as
\begin{equation}\label{Optimization_solution2}
	\hat{\varphi}_{i+1}=\min\big\{\psi_{m}\vert \psi_{m}> \varphi_{i+1}^{(p_{\star})},\ m\in\mathcal{I}_{\mathrm{M}} \big\}.
\end{equation}

\begin{algorithm}[!t]
	\caption{SDR-DC-BiS Scheme}\label{alg2}
	\begin{algorithmic}[1]
		\STATE \emph{Input:} $\varphi_{i}$, $\rho_{1}$, $\epsilon_{1}$, $\epsilon_{2}$, $\{\!\boldsymbol{A}_{m}\!\}_{m\in \!\mathcal{I}_{\mathrm{M}}}$, $\{\!\psi_m\!\}_{m\in \mathcal{I}_{\!\mathrm{M}}\!}$, $\{\!\gamma_m\!\}_{m\!\in\!\mathcal{I}_{\mathrm{M}}}$.
		\STATE Set $p\leftarrow	1$. Initialize $\varphi_{i+1}^{\mathrm{lb}}$ and $\varphi_{i+1}^{\mathrm{ub}}$.
		\WHILE {$\varphi_{i+1}^{\mathrm{ub}}-\varphi_{i+1}^{\mathrm{lb}}>\epsilon_{\varphi}$}
		\STATE Set $\varphi_{i+1}^{(p)}\leftarrow(\varphi_{i+1}^{\mathrm{lb}}+\varphi_{i+1}^{\mathrm{ub}})/2$.
		\STATE Obtain $\boldsymbol{F}_{i}^{(0)}$ through solving~\eqref{DC_SDR_Initialization}.
		\IF{$D_1(\boldsymbol{F}_{i}^{(0)})>-1$}
		\STATE Update $\varphi_{i+1}^{\mathrm{ub}}\leftarrow \varphi_{i+1}^{(p)}$.
		\ELSE
		\STATE Set $q\leftarrow 0$. 
		\REPEAT
		\STATE Obtain $\boldsymbol{\Xi}_{i}^{(q)}$ via~\eqref{Subdifferential_Rank_Matrix}.
		\STATE Obtain $\boldsymbol{F}_{i}^{(q+1)}$ through solving~\eqref{DC_Subproblem_Smooth}.
		\STATE Update $q\leftarrow q+1$.
		\UNTIL{\eqref{DC_Stop_Conditions} is satisfied}
		\STATE Obtain $\boldsymbol{f}_i^{(q)}$ via~\eqref{Optimization_solution1}.
		\IF{$\boldsymbol{f}_i^{(q)}$ is feasible for~\eqref{Feasibility}}
		\STATE Update $\varphi_{i+1}^{\mathrm{lb}}\leftarrow \varphi_{i+1}^{(p)}$ and obtain $\hat{\boldsymbol{f}}_{i}\leftarrow \boldsymbol{f}_i^{(q)}$.
		\ELSE
		\STATE Update $\varphi_{i+1}^{\mathrm{ub}}\leftarrow \varphi_{i+1}^{(p)}$.
		\ENDIF
		\ENDIF
		\STATE Update $p\leftarrow p+1$.
		\ENDWHILE 
		\STATE Obtain $\hat{\varphi}_{i+1}$ via~\eqref{Optimization_solution2}.
		\STATE \emph{Output:} $\hat{\boldsymbol{f}}_{i}$, $\hat{\varphi}_{i+1}$.
	\end{algorithmic}
\end{algorithm}

\subsection{Complexity of SDR-DC-BiS}\label{Complexity}
The proposed SDR-DC-BiS scheme is summarized in~\textbf{Algorithm~\ref{alg2}}. Now we analyze the computational complexity of SDR-DC-BS. Denote the gap between the initial upper and lower bounds of $\varphi_{i+1}$ as $\Delta\Phi_{i}$. Then, the total number of BiS can be expressed as $N_{\mathrm{P}}\triangleq\big\lceil\log_2(\Delta\Phi_{i}/\epsilon_{\psi})\big\rceil$~\cite{10520169}. During the $p$th iteration, the worst case complexity for solving~\eqref{DC_Subproblem_Smooth} or~\eqref{DC_SDR_Initialization} with the interior point method is $\mathcal{O}\big(\max\{M_{i}^{(p)},N_{\mathrm{T}}+1\}^4\sqrt{N_{\mathrm{T}}+1}\ln(1/\tilde{\epsilon}_{1})\big)$~\cite{luoSemidefiniteRelaxationQuadratic2010}, where $\tilde{\epsilon}_{1}$ represents the accuracy and $M_{i}^{(p)}$ represents the number of elements in $\mathcal{M}(\varphi_{i},\varphi_{i+1}^{(p)})$. If $\rho_{1}$ is sufficiently large, it takes fewer than $\big(D(\boldsymbol{F}_{i}^{(0)})-D(\boldsymbol{F}_{i}^{(q_\star)})\big)/\epsilon_{1}$ iterations for $\{\boldsymbol{F}_{i}^{(q)}\}$ to converge to $\boldsymbol{F}_{i}^{(q_\star)}$. Finally, the worst case complexity of SDR-DC-BiS is given by $\mathcal{O}\big(\sum_{p=1}^{N_{\mathrm{P}}}\max\{M_{i}^{(p)},N_{\mathrm{T}}+1\}^4\sqrt{N_{\mathrm{T}}+1}\ln(1/\tilde{\epsilon}_{1})/\epsilon_{1}\big)$.

In fact, the complexity of SDR-DC-BiS is high due to the two main reasons as follows. Firstly, SDR substantially increases the number of optimization variables from $N_{\mathrm{T}}$ to $N_{\mathrm{T}}^2$, which is advantageous for small $N_{\mathrm{T}}$  since the non-convex constraints in~\eqref{Feasibility} can be simplified as linear ones. However, it becomes problematic as $N_{\mathrm{T}}$ grows large, making~\eqref{DC_Subproblem} unsuitable for a large number of iterations as a basic sub-problem. Secondly, the smoothing method used in~\eqref{DC_Subproblem_Smooth} introduces a large number of constraints, significantly increasing the computational complexity associated with the interior point method~\cite{10520169}. To reduce the complexity, we will propose a scheme named PP-PDG-MS in the following section, which is better suited for cases that either $N_{\mathrm{T}}$ or $M_{i}^{(p)}$ is large.

\section{PP-PDG-MS Scheme}\label{Method1}
In this section, we propose the PP-PDG-MS scheme to solve~\eqref{Optimization} in cases that either $N_{\mathrm{T}}$ or $M_{i}^{(p)}$ is large. In the first stage, we transform~\eqref{Feasibility} into a non-convex min-max problem through relaxing the RSNR constraints and constant modulus constraints into a penalty of the objective function. The transformed problem is then solved based on a PP method and a PDG algorithm with closed-form solutions. In the second stage, a MS method is proposed, where a monotonic search is employed to establish upper and lower bounds for $\varphi_{i+1}^{\star}$, and then BiS is used to determine $\hat{\varphi}_{i+1}$. 

\subsection{Proximal-Point Method}
For the similar reason mentioned above~\eqref{Feasibility_SDR_Rank_Penalty}, with $\varphi_{i+1}$ fixed as $\varphi_{i+1}^{(p)}$,~\eqref{Feasibility} can be relaxed into
\begin{equation}\label{maxmin}
	\begin{aligned}
		\min_{\boldsymbol{f}_{i}}\max_{m\in\mathcal{M}(\varphi_{i},\varphi_{i+1}^{(p)})}\gamma_{m}-\boldsymbol{f}_{i}^{\mathrm{H}}\boldsymbol{A}_{m}\boldsymbol{f}_{i}\quad\mathrm{s.t.}~\eqref{Constant_Constraint},
	\end{aligned}
\end{equation}
where we use the difference between $\gamma_{m}$ and $\boldsymbol{f}_{i}^{\mathrm{H}}\boldsymbol{A}_{m}\boldsymbol{f}_{i}$ instead of a fraction to reduce the complexity, as will be explained in Section~\ref{complexity_PP}. Note that~\eqref{maxmin} is a non-convex non-smooth optimization problem. Solving such problems is time-consuming, even in an unconstrained setting~\cite{zhangLowerIteration2022}. Therefore, to enhance computational efficiency, we moderately relax~\eqref{Constant_Constraint} and correspondingly introduce a penalty term into the objective function of~\eqref{maxmin}.

We begin by considering the following equivalence
\begin{equation}
	\big\vert[\boldsymbol{f}_i]_{n}\big\vert\!=\!\frac{1}{\sqrt{N_{\mathrm{T}}}}\Longleftrightarrow\begin{cases}\boldsymbol{f}_i^{\mathrm{H}}\mathrm{Diag}\{\boldsymbol{e}_{n}\}\boldsymbol{f}_i\!\leq\! 1/N_{\mathrm{T}}\\ 1/N_{\mathrm{T}}-\boldsymbol{f}_i^{\mathrm{H}}\mathrm{Diag}\{\boldsymbol{e}_{n}\}\boldsymbol{f}_i\!\leq\! 0\end{cases}\!\!\!\!\!\!\!,
\end{equation} 
where $\boldsymbol{e}_{n}\in\mathbb{R}^{N_{\mathrm{T}}}$ is a vector with only one non-zero entry, i.e., $[\boldsymbol{e}_{n}]_{n}=1$. In light of this equivalence, we introduce a penalty parameter $\rho_{2}$ and formulate the penalty problem of~\eqref{maxmin} as 
\begin{subequations}\label{maxmin1}
	\begin{align}
		\min_{\boldsymbol{f}_{i}}& \max_{m\in\mathcal{M}(\varphi_{i},\varphi_{i+1}^{(p)})}\gamma_{m}-\boldsymbol{f}_{i}^{\mathrm{H}}\boldsymbol{A}_{m}\boldsymbol{f}_{i}\notag\\\
		&\quad\quad+\rho_{2}\sum_{n\in\mathcal{I}_{\mathrm{N}}} \max\Big\{\frac{1}{N_{\mathrm{T}}}-\boldsymbol{f}_i^{\mathrm{H}}\mathrm{Diag}\{\boldsymbol{e}_{n}\}\boldsymbol{f}_i,0\Big\}\label{maxmin1_Obj}\\
		\mathrm{s.t.}&\quad\boldsymbol{f}_i^{\mathrm{H}}\mathrm{Diag}\{\boldsymbol{e}_{n}\}\boldsymbol{f}_i\leq \frac{1}{N_{\mathrm{T}}},\ \forall n\in\mathcal{I}_{\mathrm{N}}.\label{maxmin1_Cons}
	\end{align}
\end{subequations}
Note that~\eqref{maxmin1_Cons} defines a convex feasible set, making~\eqref{maxmin1} much easier to solve than~\eqref{maxmin}. 

To simplify the following analysis, we transform the variables and parameters in~\eqref{maxmin1} into the real domain. Define $\tilde{\boldsymbol{f}}_{i}\triangleq\big[\mathrm{Re}\{\boldsymbol{f}_{i}\}^\mathrm{T},\mathrm{Im}\{\boldsymbol{f}_{i}\}^\mathrm{T}\big]^\mathrm{T}$ and denote the smallest element in $\mathcal{M}(\varphi_{i},\varphi_{i+1}^{(p)})$ as $m_i$. For $j=m-m_i+1$, where $m\in\mathcal{M}(\varphi_{i},\varphi_{i+1}^{(p)})$, we further define $\tilde{\gamma}_{j}\triangleq \gamma_{m}$ and 
\begin{equation}
	\tilde{\boldsymbol{A}}_{j}\triangleq\begin{bmatrix}
		\mathrm{Re}\{\boldsymbol{A}_{m}\}&-\mathrm{Im}\{\boldsymbol{A}_{m}\} \\
		\mathrm{Im}\{\boldsymbol{A}_{m}\}&\mathrm{Re}\{\boldsymbol{A}_{m}\}
		\end{bmatrix}+\rho_2\boldsymbol{I}_{2N_{\mathrm{T}}}.
\end{equation}

The objective function in~\eqref{maxmin1_Obj} can then be reformulated as
\begin{equation}\label{Original_Function}
	U(\tilde{\boldsymbol{f}}_{i})\triangleq\max_{j\in\mathcal{J}_{i}^{(p)}}\ u_{j}(\tilde{\boldsymbol{f}}_{i}),
\end{equation}
where $u_{j}(\tilde{\boldsymbol{f}}_{i})\triangleq\tilde{\gamma}_{j}-\tilde{\boldsymbol{f}}_i^{\mathrm{T}}\tilde{\boldsymbol{A}}_{j}\tilde{\boldsymbol{f}}_i$. The optimization problem in~\eqref{maxmin1} can be equivalently rewritten as
\begin{equation}\label{maxmin2}
	\min_{\tilde{\boldsymbol{f}}_{i}\in\mathcal{F}_{2}}U(\tilde{\boldsymbol{f}}_{i}),
\end{equation}
where $\mathcal{F}_{2}$ is defined as
\begin{equation}
	\mathcal{F}_{2} \triangleq\big\{\tilde{\boldsymbol{f}}_{i}\in\mathbb{R}^{2N_{\mathrm{T}}}\big\vert\tilde{\boldsymbol{f}}_i^{\mathrm{T}}\mathrm{Diag}\big\{[\boldsymbol{e}_{n}^{\mathrm{T}},\boldsymbol{e}_{n}^{\mathrm{T}}]\big\}\tilde{\boldsymbol{f}}_i\leq 1/N_{\mathrm{T}}\big\}.
\end{equation}

Solving~\eqref{maxmin2} is challenging due to the non-convexity of $u_j(\tilde{\boldsymbol{f}}_{i})$ and the non-smoothness of $U(\tilde{\boldsymbol{f}}_{i})$. Specifically, finding the global minimizer of~\eqref{maxmin2} is known to be NP-hard. Moreover, the process of verifying whether a given feasible solution $\tilde{\boldsymbol{f}}_{i}'\in \mathcal{F}_{2}$ meets the first order optimality condition incurs high computational complexity, due to the non-smoothness of $U(\tilde{\boldsymbol{f}}_{i})$~\cite{razaviyaynNonconvexMinMax2020}. Given these challenges, existing studies resort to searching for a nearly $\epsilon_{3}$-stationary point, which can be derived from a smooth function, i.e., the Moreau envelope function~\cite{davisProximallyGuided2019,davisStochasticModelBased2019,NEURIPS2019_05d0abb9}. 

In this context, the Moreau envelope function of $U(\tilde{\boldsymbol{f}}_{i})$ can be established as
\begin{equation}\label{Proximal_ME}
	U_{\mu}(\tilde{\boldsymbol{f}}_{i}) = \inf_{\boldsymbol{f}\in \mathcal{F}_{2}}\Big\{U(\boldsymbol{f})+\frac{1}{2\mu}\big\Vert\tilde{\boldsymbol{f}}_{i}-\boldsymbol{f}\big\Vert_{2}^2\Big\},
\end{equation}
where $\mu$ is a smoothing parameter. To explain the near-stationarity of $U(\tilde{\boldsymbol{f}}_{i})$ based on~\eqref{Proximal_ME}, we define $L_{i}\triangleq 2\max_{j\in\mathcal{J}_{i}^{(p)}}\big\{\rho_{\mathrm{max}}(\tilde{\boldsymbol{A}}_{j})\big\}$, where $\rho_{\mathrm{max}}(\tilde{\boldsymbol{A}}_{j})$ represents the maximum eigenvalue of $\tilde{\boldsymbol{A}}_{j}$. Note that in our specific case, we have $\rho_{\mathrm{max}}(\tilde{\boldsymbol{A}}_{j})=1+\rho_{2}$ for all $j\in\mathcal{J}_{i}^{(p)}$. Therefore, to simplify the notation, we omit the subscripts for $L_{i}$ and introduce $L=2(1+\rho_{2})$. Since $(L/2)\boldsymbol{I}_{2N_{\mathrm{T}}}-\tilde{\boldsymbol{A}}_{j}$ is positive semidefinite, $u_{j}(\tilde{\boldsymbol{f}}_{i})+(L/2)\Vert\tilde{\boldsymbol{f}}_{i}\Vert_2^{2}$ is a convex function for any $\tilde{\boldsymbol{f}}_{i}\in\mathcal{F}_{2}$, and so is $U(\tilde{\boldsymbol{f}}_{i})+(L/2)\Vert\tilde{\boldsymbol{f}}_{i}\Vert_2^{2}$. This implies that $U(\tilde{\boldsymbol{f}}_{i})$ is $L$-weakly convex. For $\mu\in(0,1/L)$, $U_{\mu}(\tilde{\boldsymbol{f}}_{i})$ is smooth and its gradient can be expressed as $\boldsymbol{\nabla}U_{\mu}\big(\tilde{\boldsymbol{f}}_{i})=\mu^{-1}(\tilde{\boldsymbol{f}}_{i}-\boldsymbol{\mathrm{prox}}_{\mu U}(\tilde{\boldsymbol{f}}_{i})\big)$~\cite{davisStochasticModelBased2019}, where $\boldsymbol{\mathrm{prox}}_{\mu U}$ is a proximal operator defined as
\begin{equation}\label{Proximal_step}
	\boldsymbol{\mathrm{prox}}_{\mu U}(\tilde{\boldsymbol{f}}_{i}) \triangleq \arg\min_{\boldsymbol{f}\in \mathcal{F}_{2}}\Big\{U(\boldsymbol{f})+\frac{1}{2\mu}\big\Vert\tilde{\boldsymbol{f}}_{i}-\boldsymbol{f}\big\Vert_{2}^2\Big\}.
\end{equation}
A feasible solution, denoted as $\tilde{\boldsymbol{f}}_{i}'$, is referred to as a nearly $\epsilon_{3}$-stationary point if it satisfies $\Vert\boldsymbol{\nabla}U_{\mu}\big(\tilde{\boldsymbol{f}}_{i}')\Vert_{2}\leq \epsilon_{3}$~\cite{NEURIPS2019_05d0abb9}. These solutions can provide good approximation for first-order stationary points if $\epsilon_{3}$ is sufficiently small~\cite{davisStochasticModelBased2019}. Furthermore, the verification process for nearly stationary points is much easier, facilitating the derivation of efficient algorithms.

The most intuitive method to search for a nearly $\epsilon_{3}$-stationary point is to iteratively set $\tilde{\boldsymbol{f}}_{i}^{(q+1)}=\boldsymbol{\mathrm{prox}}_{\mu U}(\tilde{\boldsymbol{f}}_{i}^{(q)})$. However, this method is not considered in this paper due to its high computational complexity. Specifically, the problem in~\eqref{Proximal_step}  lacks closed-form solutions, necessitating its decomposition into sub-problems that must be iteratively solved. To avoid additional iterative loops, we establish a surrogate function for the objective function in~\eqref{Proximal_step} as
\begin{equation}\label{Proximal_Linear}
	\begin{aligned}
		\hat{U}\big(\tilde{\boldsymbol{f}}_{i},\tilde{\boldsymbol{f}}_{i}^{(q)}\big)\!\triangleq\!\max_{j\in\mathcal{J}_{i}^{(p)}}& u_{j}\big(\tilde{\boldsymbol{f}}_{i}^{(q)}\big)\!+\!\boldsymbol{\nabla}^{\mathrm{T}}\!u_{j}(\tilde{\boldsymbol{f}}_{i}^{(q)})\!\big(\tilde{\boldsymbol{f}}_{i}\!-\!\tilde{\boldsymbol{f}}_{i}^{(q)}\big)\! \\
		&+ \frac{\sigma}{2}\big\Vert \tilde{\boldsymbol{f}}_{i}-\tilde{\boldsymbol{f}}_{i}^{(q)}\big\Vert_{2}^{2},\\
	\end{aligned}
\end{equation}
where $\sigma\triangleq\mu^{-1}-L$ is a parameter introduced to ensure the strong convexity of $\hat{U}\big(\tilde{\boldsymbol{f}}_{i},\tilde{\boldsymbol{f}}_{i}^{(q)}\big)$. Note that $\hat{U}\big(\tilde{\boldsymbol{f}}_{i},\tilde{\boldsymbol{f}}_{i}^{(q)}\big)$ provides an upper bound for $U(\tilde{\boldsymbol{f}}_{i}^{(q)})$ and a lower bound for $U(\tilde{\boldsymbol{f}}_{i}^{(q)})+(\mu^{-1}/2)\big\Vert\tilde{\boldsymbol{f}}_{i}-\tilde{\boldsymbol{f}}_{i}^{(q)}\big\Vert_{2}^2$ . Specifically, since $(L/2)\boldsymbol{I}_{2N_{\mathrm{T}}}-\tilde{\boldsymbol{A}}_{j}$ is positive semidefinite, we have
\begin{equation}\label{Proxi_prove3}
	\begin{aligned}		
		&u_{j}(\tilde{\boldsymbol{f}}_{i})\!\geq\! u_{j}(\tilde{\boldsymbol{f}}_{i}^{(q)})\!+\!\boldsymbol{\nabla}^{\mathrm{T}}u_{j}(\tilde{\boldsymbol{f}}_{i}^{(q)})(\tilde{\boldsymbol{f}}_{i}\!-\!\tilde{\boldsymbol{f}}_{i}^{(q)})\!-\!\frac{L}{2}\Vert\tilde{\boldsymbol{f}}_{i}\!-\!\tilde{\boldsymbol{f}}_{i}^{(q)}\Vert_{2}^{2}\\
		&\Longrightarrow U(\tilde{\boldsymbol{f}}_{i})+\frac{\mu^{-1}}{2}\Vert\tilde{\boldsymbol{f}}_{i}-\tilde{\boldsymbol{f}}_{i}^{(q)}\Vert_{2}^{2}\geq\hat{U}\big(\tilde{\boldsymbol{f}}_{i},\tilde{\boldsymbol{f}}_{i}^{(q)}\big).
	\end{aligned}
\end{equation}
Moreover, since $\tilde{\boldsymbol{A}}_{j}\!-\!\rho_{2}\boldsymbol{I}_{2N_{\mathrm{T}}}$ is semidefinite positive, we have
\begin{equation}\label{Proxi_prove4}
	\begin{aligned}	
		&u_{j}(\tilde{\boldsymbol{f}}_{i}^{(q)})\!+\!\boldsymbol{\nabla}^{\mathrm{T}}u_{j}(\tilde{\boldsymbol{f}}_{i}^{(q)})(\tilde{\boldsymbol{f}}_{i}\!-\!\tilde{\boldsymbol{f}}_{i}^{(q)})\!\geq\! u_{j}(\tilde{\boldsymbol{f}}_{i})\!+\!\rho_2\Vert\tilde{\boldsymbol{f}}_{i}\!-\!\tilde{\boldsymbol{f}}_{i}^{(q)}\Vert_{2}^{2}\\
	&\Longrightarrow\hat{U}\big(\tilde{\boldsymbol{f}}_{i},\tilde{\boldsymbol{f}}_{i}^{(q)}\big)\geq U(\tilde{\boldsymbol{f}}_{i})+\frac{\sigma+2\rho_2}{2}\Vert\tilde{\boldsymbol{f}}_{i}-\tilde{\boldsymbol{f}}_{i}^{(q)}\Vert_{2}^{2}.
	\end{aligned}
\end{equation} 

We then generate a sequence $\{\tilde{\boldsymbol{f}}_{i}^{(q)}\}$ by iteratively finding a solution $\tilde{\boldsymbol{f}}_{i}^{(q+1)}\in\mathcal{F}_{2}$ such that
\begin{equation}\label{Proximal_require}
	\hat{U}\big(\tilde{\boldsymbol{f}}_{i}^{(q+1)},\tilde{\boldsymbol{f}}_{i}^{(q)}\big)\leq\hat{U}\big(\tilde{\boldsymbol{f}}_{i}^{(q,\star)},\tilde{\boldsymbol{f}}_{i}^{(q)}\big)+\tilde{\epsilon}_{2},
\end{equation}
where $\tilde{\boldsymbol{f}}_{i}^{(q,\star)}\triangleq\arg\min_{\tilde{\boldsymbol{f}}_{i}\in\mathcal{F}_{2}}\hat{U}\big(\tilde{\boldsymbol{f}}_{i},\tilde{\boldsymbol{f}}_{i}^{(q)}\big)$, and $\tilde{\epsilon}_{2}$ is a predetermined error tolerance threshold. The iterations stop once
\begin{equation}\label{Proximal_error_require}
	U\big(\tilde{\boldsymbol{f}}_{i}^{(q)}\big)-\hat{U}\big(\tilde{\boldsymbol{f}}_{i}^{(q+1)},\tilde{\boldsymbol{f}}_{i}^{(q)}\big) \leq \epsilon_{3}^2\mu^2\sigma/8-\tilde{\epsilon}_{2},
\end{equation}
where $\tilde{\epsilon}_{2}<\epsilon_{3}^2\mu^2\sigma/8$~\cite{NEURIPS2019_05d0abb9}. If $\tilde{\epsilon}_{2}$ is sufficiently small,~\eqref{Proximal_error_require} can be satisfied within a limit number of iterations, and the obtained solution, denoted as $\tilde{\boldsymbol{f}}_{i}^{(q_\star)}$, is guaranteed to be nearly $\epsilon_{3}$-stationary. To prove it, we introduce the inequalities
\begin{equation}\label{Proxi_prove1}
	\begin{aligned}
		U\big(\tilde{\boldsymbol{f}}_{i}^{(q)}\big)&\!=\!\hat{U}\big(\tilde{\boldsymbol{f}}_{i}^{(q)},\tilde{\boldsymbol{f}}_{i}^{(q)}\big)\\
		&\!\geq\! \hat{U}\big(\tilde{\boldsymbol{f}}_{i}^{(q,\star)},\tilde{\boldsymbol{f}}_{i}^{(q)}\big)+\frac{\sigma}{2}\big\Vert\tilde{\boldsymbol{f}}_{i}^{(q,\star)} -\tilde{\boldsymbol{f}}_{i}^{(q)}\big\Vert_{2}^{2}\\
		&\!\geq\! \hat{U}\big(\tilde{\boldsymbol{f}}_{i}^{(q+1)},\tilde{\boldsymbol{f}}_{i}^{(q)}\big)\!+\!\frac{\sigma}{2}\big\Vert\tilde{\boldsymbol{f}}_{i}^{(q,\star)} -\tilde{\boldsymbol{f}}_{i}^{(q)}\big\Vert_{2}^{2} \!-\!\tilde{\epsilon}_{2},
	\end{aligned}
\end{equation}
\begin{equation}\label{Proxi_prove2}
	\hat{U}\!\big(\tilde{\boldsymbol{f}}_{i}^{(q+1)}\!,\!\tilde{\boldsymbol{f}}_{i}^{(q)}\big)\!\geq\! U\!\big(\tilde{\boldsymbol{f}}_{i}^{(q+1)}\big)\!+\frac{\sigma\!+\!2\rho_{2}}{2}\big\Vert\tilde{\boldsymbol{f}}_{i}^{(q+1)}\!-\!\tilde{\boldsymbol{f}}_{i}^{(q)}\big\Vert_{2}^{2},
\end{equation}
where~\eqref{Proxi_prove1} holds since $\hat{U}\big(\tilde{\boldsymbol{f}}_{i},\tilde{\boldsymbol{f}}_{i}^{(q)}\big)$ is $\sigma$-strongly convex, and~\eqref{Proxi_prove2} holds due to~\eqref{Proxi_prove4}. To simplify the notation, we temporarily set $\boldsymbol{p}\triangleq\boldsymbol{\mathrm{prox}_{\mu U}}(\tilde{\boldsymbol{f}}_{i}^{(q)})$, $\boldsymbol{f}\triangleq\tilde{\boldsymbol{f}}_{i}^{(q)}$ and $\boldsymbol{f}_{\star}\triangleq\tilde{\boldsymbol{f}}^{(q,\star)}$. If~\eqref{Proximal_error_require} is satisfied at the $q$th iteration, we have $\big\Vert\boldsymbol{f}_{\star} -\boldsymbol{f}\big\Vert_{2}^{2}\leq(\epsilon_{3}^2\mu^2)/4$ according to~\eqref{Proxi_prove1}, which leads to
\begin{subequations}
	\begin{align}
		&\Vert \boldsymbol{\nabla}U_{\mu_{i}}(\boldsymbol{f})\Vert_{2}^{2}=\mu^{-2}\Vert\boldsymbol{f}-\boldsymbol{p}\Vert_{2}^{2}\\
		&\!\geq\! 2\mu^{-1}\big(\hat{U}(\boldsymbol{p},\boldsymbol{f})-U(\boldsymbol{p})\big)\label{Inequality0} \\
		&\!\geq\! 2\mu^{-1}\Big(\hat{U}(\boldsymbol{f}_\star,\boldsymbol{f})+\frac{\sigma}{2}\Vert\boldsymbol{f}_\star-\boldsymbol{p}\Vert_{2}^{2}\Big)-2\mu^{-1}U(\boldsymbol{p})\label{Inequality1} \\
		&\!\geq\! 2\mu^{-1}\Big(U(\boldsymbol{f})\!-\!\frac{\epsilon_{3}^2\mu^2\sigma}{8}\!+\!\frac{\sigma}{2}\Vert\boldsymbol{f}_\star\!-\!\boldsymbol{p}\Vert_{2}^{2}\Big)\!-\! 2\mu^{-1}U(\boldsymbol{p}) \label{Inequality2}\\
		&\!\geq\! \mu^{-2}\Vert\boldsymbol{f}-\boldsymbol{p}\Vert_{2}^{2}- \frac{\epsilon_{3}^2\mu \sigma}{4}+\mu^{-1}\sigma\Vert\boldsymbol{f}_\star-\boldsymbol{p}\Vert_{2}^{2}\label{Inequality3}\\
		&\Longrightarrow  \big\vert\Vert\boldsymbol{f}_\star-\boldsymbol{f}\Vert_{2}-\Vert\boldsymbol{f}-\boldsymbol{p}\Vert_{2}  \big \vert \leq\Vert\boldsymbol{f}_\star-\boldsymbol{p}\Vert_{2}\leq \frac{\epsilon_{3}\mu}{2}\label{Inequality4}\\
		&\Longrightarrow \Vert \boldsymbol{\nabla}U_{\mu_{i}}(\boldsymbol{f})\Vert_{2}\leq \frac{\epsilon_{3}}{2}+\mu^{-1}\Vert\boldsymbol{f}-\boldsymbol{f}_{\star}\Vert_{2} \leq \epsilon_{3}.
	\end{align}
\end{subequations}
Note that~\eqref{Inequality1} holds since $\hat{U}\big(\tilde{\boldsymbol{f}}_{i},\tilde{\boldsymbol{f}}_{i}^{(q)}\big)$ is $\sigma$-strongly convex,~\eqref{Inequality3} holds since $U(\tilde{\boldsymbol{f}}_{i}^{(q)})\geq U(\boldsymbol{\mathrm{prox}_{\mu U}}\big(\tilde{\boldsymbol{f}}_{i}^{(q)})\big)+(\mu^{-1}/2)\big\Vert\tilde{\boldsymbol{f}}_{i}^{(q)}-\boldsymbol{\mathrm{prox}_{\mu U}}(\tilde{\boldsymbol{f}}_{i}^{(q)})\big\Vert_{2}^2$, and~\eqref{Inequality4} is derived using the triangle inequality. If~\eqref{Proximal_error_require} is not satisfied at the $q$th iteration,~\eqref{Proxi_prove2} indicates that $\{U\big(\tilde{\boldsymbol{f}}_{i}^{(q)}\big)\}$ is decreasing. Furthermore, since $U(\tilde{\boldsymbol{f}}_{i})$ is lower bounded and $\tilde{\epsilon}_{2}$ is sufficiently small,~\eqref{Proxi_prove1} and~\eqref{Proxi_prove2} indicate that as $q$ becomes large enough, $U(\tilde{\boldsymbol{f}}_{i}^{(q)})-\hat{U}(\tilde{\boldsymbol{f}}_{i}^{(q+1)},\tilde{\boldsymbol{f}}_{i}^{(q)})$ will approach zero and finally satisfy~\eqref{Proximal_error_require}.

\subsection{Primal-Dual Gradient Algorithm}\label{PDG}
To efficiently find a solution $\tilde{\boldsymbol{f}}_{i}^{(q+1)}\in\mathcal{F}_{2}$ that satisfies~\eqref{Proximal_require}, the non-smoothness of $\hat{U}\big(\tilde{\boldsymbol{f}}_{i},\tilde{\boldsymbol{f}}_{i}^{(q)}\big)$ should be addressed. To this end, we introduce an auxiliary vector $\boldsymbol{z}_{i}\in \mathbb{R}^{M_{i}^{(p)}}$ and reformulate $\hat{U}\big(\tilde{\boldsymbol{f}}_{i},\tilde{\boldsymbol{f}}_{i}^{(q)}\big)$ as
\begin{equation}\label{Proximal_Linear1}
		\begin{aligned}
			\!\hat{U}\!\big(\tilde{\boldsymbol{f}}_{i},\!\tilde{\boldsymbol{f}}_{i}^{(q)}\big)\!=&\frac{\sigma}{2}\big\Vert \tilde{\boldsymbol{f}}_{i}\!-\!\tilde{\boldsymbol{f}}_{i}^{(q)}\big\Vert_{2}^{2}\!+\! \max_{\boldsymbol{z}_{i}\in\mathcal{Z}}\sum_{j\in\! \mathcal{J}_{i}^{(p)}\!}\! [\boldsymbol{z}_{i}]_{j}\big(u_{j}(\tilde{\boldsymbol{f}}_{i}^{(q)})\\
		&\quad\quad+\boldsymbol{\nabla}^{\mathrm{T}}u_{j}(\tilde{\boldsymbol{f}}_{i}^{(q)})(\tilde{\boldsymbol{f}}_{i}-\tilde{\boldsymbol{f}}_{i}^{(q)}) \big),\\
	\end{aligned}
\end{equation}
where $\mathcal{Z}\triangleq\big\{\boldsymbol{z}\in \mathbb{R}^{M_{i}^{(p)}}\big\vert \boldsymbol{z}\geq \boldsymbol{0}, \boldsymbol{1}^\mathrm{T}_{M_{i}^{(p)}}\boldsymbol{z}=1\big\}$. The functions in~\eqref{Proximal_Linear1} and~\eqref{Proximal_Linear} are equivalent because we have $\max_{\boldsymbol{z}\in\mathcal{Z}}\sum_{j\in \mathcal{J}_{i}^{(p)}}[\boldsymbol{z}]_{j}h_{j}(\boldsymbol{f})=\max_{j\in \mathcal{J}_{i}^{(p)}}h_{j}(\boldsymbol{f})$ for any $\boldsymbol{f}\in\mathbb{R}^{2N_\mathrm{T}}$, where $h_{j}(\boldsymbol{f})$ is an arbitrary function. This transformation is widely used in existing studies on the min-max optimization problems~\cite{NEURIPS2019_05d0abb9,nesterovExcessiveGap2005,luHybridBlock2020a}, with more detailed discussions available in~\cite{nesterovSmoothminimization2005a}.

To simplify the notation, we define $\boldsymbol{g}_{i,j}^{(q)} \triangleq -2\tilde{\boldsymbol{A}}_{j}\tilde{\boldsymbol{f}}_{i}^{(q)}$, $d_{i,j}^{(q)} \triangleq \tilde{\gamma}_{j}+\big(\tilde{\boldsymbol{f}}_{i}^{(q)}\big)^\mathrm{T}\tilde{\boldsymbol{A}}_{j}\tilde{\boldsymbol{f}}_{i}^{(q)}$, $\boldsymbol{G}^{(q)}_{i} \triangleq \big[\boldsymbol{g}_{i,1}^{(q)},\boldsymbol{g}_{i,2}^{(q)},\dots,\boldsymbol{g}_{i,M_{i}^{(p)}}^{(q)}\big]$ and $\boldsymbol{d}_{i}^{(q)} \triangleq \big[d_{i,1}^{(q)},d_{i,2}^{(q)},\dots,d_{i,M_{i}^{(p)}}^{(q)}\big]^\mathrm{T}$. The problem of minimizing $\hat{U}\big(\tilde{\boldsymbol{f}}_{i},\tilde{\boldsymbol{f}}_{i}^{(q)}\big)$ can then be rewritten as
\begin{equation}\label{Proximal_Linear2}
	\begin{aligned}
		\min_{\tilde{\boldsymbol{f}}_{i}\in\mathcal{F}_{2}} \max_{\boldsymbol{z}_{i}\in \mathcal{Z}}\quad&\mathcal{U}^{(q)}\big(\tilde{\boldsymbol{f}}_{i},\boldsymbol{z}_{i}\big)\triangleq\frac{\sigma}{2}\big\Vert \tilde{\boldsymbol{f}}_{i}-\tilde{\boldsymbol{f}}_{i}^{(q)}\big\Vert_{2}^{2}\\
		&\quad\quad\quad+\tilde{\boldsymbol{f}}_{i}^\mathrm{T}\boldsymbol{G}_{i}^{(q)}\boldsymbol{z}_{i}+\big(\boldsymbol{d}_{i}^{(q)}\big)^\mathrm{T}\boldsymbol{z}_{i}.
	\end{aligned}
\end{equation}
Note that~\eqref{Proximal_Linear2} is a strongly-convex-concave min-max problem where $\tilde{\boldsymbol{f}}_{i}$ and $\boldsymbol{z}_{i}$ are associated with a bilinear function. In this paper, we adopt the PDG method to solve~\eqref{Proximal_Linear2}~\cite{nesterovExcessiveGap2005}. We start by introducing a distance generating function $d(\boldsymbol{z}_{i})$ in $\mathcal{U}^{(q)}\big(\tilde{\boldsymbol{f}}_{i},\boldsymbol{z}_{i}\big)$ to construct an approximate function of $\hat{U}\!\big(\tilde{\boldsymbol{f}}_{i},\tilde{\boldsymbol{f}}_{i}^{(q)}\big)$ as
\begin{equation}\label{Proximal_Approx}
	\hat{U}_{\mu_{\mathrm{z}}}\big(\tilde{\boldsymbol{f}}_{i},\tilde{\boldsymbol{f}}_{i}^{(q)}\big)=\max_{\boldsymbol{z}_{i}\in \mathcal{Z}}\mathcal{U}_{\mu_{\mathrm{z}}}^{(q)}\big(\tilde{\boldsymbol{f}}_{i},\boldsymbol{z}_{i}\big),
\end{equation}
where $\mu_{\mathrm{z}}$ is a positive variable introduced to control the approximation error, and
\begin{equation}\label{Proximal_SConcave}
	\mathcal{U}_{\mu_{\mathrm{z}}}^{(q)}\big(\tilde{\boldsymbol{f}}_{i},\boldsymbol{z}_{i}\big) = \mathcal{U}^{(q)}\big(\tilde{\boldsymbol{f}}_{i},\boldsymbol{z}_{i}\big)-\mu_{\mathrm{z}}d(\boldsymbol{z}_{i}).
\end{equation}
Note that $d(\boldsymbol{z}_{i})$ should be strongly convex, and satisfy the following two requirements: (i) $d(\boldsymbol{z}_{i})\geq (1/2)\Vert \boldsymbol{z}_{i}-\boldsymbol{z}_{0}\Vert^{2}$ holds for any $\boldsymbol{z}_{i}\in\mathcal{Z}$, where $\Vert\cdot\Vert$ is an arbitrary norm; (ii) $d(\boldsymbol{z}_{0})=\min_{\boldsymbol{z}_{i}\in\mathcal{Z}}d(\boldsymbol{z}_{i})=0$. The incorporation of $d_2(\boldsymbol{z}_{i})$ ensures that $\mathcal{U}_{\mu_{\mathrm{z}}}^{(q)}\big(\tilde{\boldsymbol{f}}_{i},\boldsymbol{z}_{i}\big)$ is strongly concave with respect to $\boldsymbol{z}_{i}$, leading to a uniquely defined optimal solution for the problem in~\eqref{Proximal_Approx}. According to the Danskin’s theorem, since the optimal solution set of the problem in~\eqref{Proximal_Approx} contains only one element, $\hat{U}_{\mu_{\mathrm{z}}}\big(\tilde{\boldsymbol{f}}_{i},\tilde{\boldsymbol{f}}_{i}^{(q)}\big)$ is differentiable. The uniqueness of the optimal solution and the differentiability of $\hat{U}_{\mu_{\mathrm{z}}}\big(\tilde{\boldsymbol{f}}_{i},\tilde{\boldsymbol{f}}_{i}^{(q)}\big)$ enable the development of the PDG algorithm.

Each PDG iteration involves solving an auxiliary sub-problem associated with the $\ell_2$-norm, an auxiliary sub-problem associated with $d(\boldsymbol{z}_{i})$, and a gradient mapping sub-problem associated with the norm specified in requirement~(i). Consequently, the efficiency of PDG is significantly dependent on the formulation of $d(\boldsymbol{z}_{i})$. In this paper, different from~\cite{nesterovExcessiveGap2005}, we do not use the entropy distance function, $d_{1}(\boldsymbol{z}_i)=\ln M_{i}^{(p)}+\sum_{j\in\mathcal{J}_{i}^{(p)}}[\boldsymbol{z}_i]_{j}\ln [\boldsymbol{z}_i]_{j},\ \boldsymbol{z}_i\in\mathcal{Z}$ and its associated $\ell_{1}$-norm. This is primarily due to potential ill-conditioning issues involved in solving the auxiliary sub-problem associated with $d_{1}(\boldsymbol{z}_i)$, especially when $\mu_{\mathrm{z}}$ is small. These issues arise due to the involvement of exponential components in the solution of the sub-problem, such as $\exp\big(\mu_{\mathrm{z}}^{-1}(\boldsymbol{g}_{i,j}^{(q)})^{\mathrm{T}}\tilde{\boldsymbol{f}}_{i}+\mu_{\mathrm{z}}^{-1}d_{i,j}^{(q)}\big)$. Furthermore, the complexity of solving the gradient mapping sub-problem associated with the $\ell_{1}$-norm is approximately $\mathcal{O}(M_{i}^{(p)}\ln M_{i}^{(p)})$, which is not advantageous in our case since $M_{i}^{(p)}$ may be large. Therefore, we  employ the traditional Euclidean distance function $d_2(\boldsymbol{z}_{i})\triangleq (1/2)\Vert \boldsymbol{z}_{i}-\boldsymbol{z}_{0}\Vert_{2}^{2}$, where $[\boldsymbol{z}_{0}]_j\triangleq 1/M_{i}^{(p)},\ \forall j\in\mathcal{J}_{i}^{(p)}$. With $d_2(\boldsymbol{z}_{i})$, the solution to each sub-problem can be derived from the Karush-Kuhn-Tucker (KKT) conditions and expressed in the closed forms as follows.

\begin{enumerate}[labelsep = .5em, leftmargin = 0em, itemindent = 1.3em]
\item{Given $\boldsymbol{z}_i$, the first auxiliary sub-problem can be given by}
\begin{equation}\label{PDG_sub1}
	\min_{\tilde{\boldsymbol{f}}_{i}\in \mathcal{F}_{2}}\  \tilde{\boldsymbol{f}}_{i}^\mathrm{T}\boldsymbol{G}_{i}^{(q)}\boldsymbol{z}_{i}+\frac{\sigma}{2}\big\Vert \tilde{\boldsymbol{f}}_{i}-\tilde{\boldsymbol{f}}_{i}^{(q)}\big\Vert_{2}^{2}.
\end{equation}
The solution of~\eqref{PDG_sub1} is unique and can be expressed in the closed form as
\begin{equation}\label{PDG_sub1_solution}
	\bar{\boldsymbol{f}}_{i}^{(q)}(\boldsymbol{z}_{i})=- \mathrm{Diag}\big\{\boldsymbol{b}(\boldsymbol{\lambda}_1)\big\}\boldsymbol{b}_{i}^{(q)},
\end{equation}
where $\boldsymbol{\lambda}_1$ is the vector of Lagrange multipliers, $[\boldsymbol{b}(\boldsymbol{\lambda}_1)]_{n}=[\boldsymbol{b}(\boldsymbol{\lambda}_1)]_{n+N_{\mathrm{T}}}=(\sigma+2[\boldsymbol{\lambda}_1]_{n})^{-1}$ for any $n\in\mathcal{I}_{\mathrm{N}}$, and $\boldsymbol{b}_{i}^{(q)}=\boldsymbol{G}_{i}^{(q)}\boldsymbol{z}_{i}\!-\!\sigma\tilde{\boldsymbol{f}}_{i}^{(q)}$. Specifically, $\boldsymbol{\lambda}_1$ can be given by
\begin{equation}
	[\boldsymbol{\lambda}_1]_{n}=\frac{1}{2}\Big(\sqrt{ N_{\mathrm{T}}\big(\big[\boldsymbol{b}_{i}^{(q)}\big]_{n}^2+\big[\boldsymbol{b}_{i}^{(q)}\big]_{n+N_{\mathrm{T}}}^2\big)}-\sigma\Big)^{+}\!,\ \forall n\in\mathcal{I}_{\mathrm{N}}. 
\end{equation}
\item{Given $\tilde{\boldsymbol{f}}_i$ and $\mu_{\mathrm{z}}$, the second auxiliary sub-problem can be given by}
\begin{equation}\label{PDG_sub2}
	\begin{aligned}
		\max_{\boldsymbol{z}_{i}\in \mathcal{Z}}\  \tilde{\boldsymbol{f}}_{i}^\mathrm{T}\boldsymbol{G}_{i}^{(q)}\boldsymbol{z}_{i}+\big(\boldsymbol{d}_{i}^{(q)}\big)^\mathrm{T}\boldsymbol{z}_{i}-\frac{\mu_{\mathrm{z}}}{2}\Vert \boldsymbol{z}_{i}-\boldsymbol{z}_{0}\Vert_{2}^{2}.
	\end{aligned}
\end{equation}
The solution of~\eqref{PDG_sub2} can be expressed as
\begin{equation}\label{PDG_sub2_solution}
	\big[\bar{\boldsymbol{z}}_{i}^{(q)}(\tilde{\boldsymbol{f}}_{i},\mu_{\mathrm{z}})\big]_{j}=\frac{1}{\mu_{\mathrm{z}}}\Big(\big[\boldsymbol{c}_{i}^{(q)}\big]_{j}-\nu_{2} \Big)^{+}\!,\ \forall j\in\mathcal{J}_{i}^{(p)},
\end{equation}
where $\boldsymbol{c}_{i}^{(q)}=\boldsymbol{d}_{i}^{(q)}+\big(\boldsymbol{G}_{i}^{(q)}\big)^{\mathrm{T}}\tilde{\boldsymbol{f}}_{i}+\mu_{\mathrm{z}}\boldsymbol{z}_{0}$, and $\nu_{2}$ should satisfy $\sum_{j\in\mathcal{J}_{i}^{(p)}}\big([\boldsymbol{c}_{i}^{(q)}]_{j}-\nu_{2} \big)^{+}=\mu_{\mathrm{z}}$. To determine the value of $\nu_{2}$, we first sort the entries of $\boldsymbol{c}_{i}^{(q)}$ in the descending order, i.e., $[\boldsymbol{c}_{i}^{(q)}]_{j_{l}}\geq[\boldsymbol{c}_{i}^{(q)}]_{j_{l+1}}$ for $l=1,2,\dots,M_{i}^{(p)}-1$, and then construct a new vector $\tilde{\boldsymbol{c}}_{i}^{(q)}$ with each entry given by $[\tilde{\boldsymbol{c}}_{i}^{(q)}]_{l}=[\boldsymbol{c}_{i}^{(q)}]_{j_{l}}$ for $l\in\mathcal{J}_{i}^{(p)}$. Then, $\nu_{2}$ can be expressed as
\begin{equation}\label{PDG_sub2_solution2}
	\nu_{2} =\frac{1}{l^{\star}}\sum_{l=1}^{l^{\star}}[\tilde{\boldsymbol{c}}_{i}^{(q)}]_{l}-\frac{1}{l^{\star}}\mu_{\mathrm{z}},
\end{equation}
where we have 
\begin{equation}
	\begin{aligned}
		l^{\star}\!=\!\max\bigg\{\!&l\in\mathcal{J}_{i}^{(p)}\bigg\vert \big[\tilde{\boldsymbol{c}}_{i}^{(q)}\big]_{l}\!>\! \frac{\sum_{l'=1}^{l}\big[\tilde{\boldsymbol{c}}_{i}^{(q)}\big]_{l'}-\mu_{\mathrm{z}}}{l} \!\bigg\}.
	\end{aligned}
\end{equation}

\item{Given $\boldsymbol{z}_i$, we solve the gradient mapping sub-problem}
\begin{equation}\label{PDG_sub3} 
		\max_{\boldsymbol{v}\in\mathcal{Z}}\Big(\!\boldsymbol{d}_{i}^{(q)}\!+ \big(\!\boldsymbol{G}_{i}^{(q)}\big)^{\mathrm{T}}\!\bar{\boldsymbol{f}}_{i}^{(q)}(\boldsymbol{z}_{i})\!\Big)^\mathrm{T}\!(\boldsymbol{v}-\boldsymbol{z}_{i})-\frac{L_{i}^{(q)}}{2}\Vert \boldsymbol{v}-\boldsymbol{z}_{i} \Vert_{2}^{2},
\end{equation}
where $L_{i}^{(q)}\triangleq \big\Vert\boldsymbol{G}_{i}^{(q)}\big\Vert_{2}^{2}/\sigma$. The solution can be given by 
\begin{equation}\label{PDG_sub3_solution}
	\big[\boldsymbol{v}_{i}^{(q)}(\boldsymbol{z}_{i})\big]_{j}\!=\!\frac{1}{L_{i}^{(q)}}\Big(\big[\boldsymbol{w}_{i}^{(q)}\big]_{j}-\nu_{3} \Big)^{+}\!,\ \forall j\in\mathcal{J}_{i}^{(p)},
\end{equation}
where $\boldsymbol{w}_{i}^{(q)}=\boldsymbol{d}_{i}^{(q)}+\big(\boldsymbol{G}_{i}^{(q)}\big)^{\mathrm{T}}\bar{\boldsymbol{f}}_{i}^{(q)}(\boldsymbol{z}_{i})+L_{i}^{(q)}\boldsymbol{z}_{i}$. The value of $\nu_{3}$ can be determined using the same method as presented in~\eqref{PDG_sub2_solution2}.
\end{enumerate}

\begin{algorithm}[t]
	\caption{PP-PDG Algorithm}\label{alg:alg3}
	\begin{algorithmic}[1]
		\STATE \emph{Input:} $\tilde{\boldsymbol{f}}_{i}^{(0)}$, $\varphi_{i+1}^{(p)}$, $w$, $\epsilon_{3}$, $\{\gamma_m\}_{m\in \mathcal{I}_{\mathrm{M}}}$, $\{\boldsymbol{A}_{m}\}_{m\in \mathcal{I}_{\mathrm{M}}}$, $\rho_{2}$, $w_{\mu}$, $Q$, $\boldsymbol{z}_{0}$
		\STATE Set $q\leftarrow -1$, $L\leftarrow 2+2\rho_{2}$, $\sigma\leftarrow (w_\mu^{-1}-1)L$ and $\tilde{\epsilon}_{2}\leftarrow ww_\mu(1-w_\mu)\epsilon_{3}^2/(8L)$.
		\REPEAT
		\STATE Update $q\leftarrow q+1$. Calculate $\boldsymbol{G}^{(q)}_{i}$, $\boldsymbol{d}^{(q)}_{i}$ and $L_{i}^{(q)}$.
		\STATE Initialize $\bar{\boldsymbol{f}}\leftarrow \bar{\boldsymbol{f}}_{i}^{(q)}(\boldsymbol{z}_{0})$ and $\bar{\boldsymbol{z}}\leftarrow \boldsymbol{v}_{i}^{(q)}(\boldsymbol{z}_{0})$.
		\STATE Initialize $\mu_{\mathrm{z}}\leftarrow 2L_i^{(q)}$ and set $k\leftarrow 0$.
		\WHILE {$\hat{U}\big(\bar{\boldsymbol{f}},\tilde{\boldsymbol{f}}_{i}^{(q)}\big)-\mathcal{U}^{(q)}\big(\bar{\boldsymbol{f}}_{i}^{(q)}(\bar{\boldsymbol{z}}),\bar{\boldsymbol{z}}\big)>\tilde{\epsilon}_{2}$}
		\STATE Set $\theta\leftarrow \frac{2}{k+3}$ and $\hat{\boldsymbol{z}}\leftarrow (1-\theta)\bar{\boldsymbol{z}}+\theta\bar{\boldsymbol{z}}_{i}^{(q)}(\bar{\boldsymbol{f}},\mu_{\mathrm{z}})$.
		\STATE Update $\mu_{\mathrm{z}}\leftarrow (1-\theta)\mu_{\mathrm{z}}$, $\bar{\boldsymbol{f}}\leftarrow(1-\theta)\bar{\boldsymbol{f}}+\theta\bar{\boldsymbol{f}}_{i}^{(q)}(\hat{\boldsymbol{z}})$ and $\bar{\boldsymbol{z}}\leftarrow\boldsymbol{v}_{i}^{(q)}(\hat{\boldsymbol{z}})$. Update $k\leftarrow k+1$.
		\ENDWHILE
		\STATE Obtain $\tilde{\boldsymbol{f}}_{i}^{(q+1)}\leftarrow \bar{\boldsymbol{f}}$. 
		\UNTIL {\eqref{Proximal_error_require} holds \OR $q\geq Q$}
		\STATE Obtain $q_\star\leftarrow q$ and $\tilde{\boldsymbol{f}}_{i}^{(q_\star)}\leftarrow \tilde{\boldsymbol{f}}_{i}^{(q)}$.
		\STATE \emph{Output:} $\tilde{\boldsymbol{f}}_{i}^{(q_\star)}$, $q_\star$.
	\end{algorithmic}
	\label{alg3}
\end{algorithm}

In each PDG iteration, the motivation of solving the aforementioned three sub-problems is to construct a primal-dual pair $(\bar{\boldsymbol{f}},\bar{\boldsymbol{z}})$ such that 
\begin{equation}\label{PDG_proof1}
	\max_{\boldsymbol{z}_{i}\in\mathcal{Z}}\ \mathcal{U}_{\mu_{\mathrm{z}}}^{(q)}\big(\bar{\boldsymbol{f}},\boldsymbol{z}_{i}\big)\leq \min_{\tilde{\boldsymbol{f}}_{i}\in \mathcal{F}_{2}} \mathcal{U}^{(q)}(\tilde{\boldsymbol{f}}_{i},\bar{\boldsymbol{z}}).
\end{equation}
Note that~\eqref{PDG_proof1} leads to a gap between $\hat{U}\big(\bar{\boldsymbol{f}}_{i},\tilde{\boldsymbol{f}}_{i}^{(q)}\big)$ and $\hat{U}\big(\tilde{\boldsymbol{f}}_{i}^{(q,\star)},\tilde{\boldsymbol{f}}_{i}^{(q)}\big)$, expressed as
\begin{equation}\label{Saddle}
	\begin{aligned}
		&\ \hat{U}\big(\bar{\boldsymbol{f}}_{i},\tilde{\boldsymbol{f}}_{i}^{(q)}\big)-\hat{U}\big(\tilde{\boldsymbol{f}}_{i}^{(q,\star)},\tilde{\boldsymbol{f}}_{i}^{(q)}\big)\\
		&=\max_{\boldsymbol{z}_{i}\in\mathcal{Z}}\ \mathcal{U}^{(q)}\big(\bar{\boldsymbol{f}},\boldsymbol{z}_{i}\big)-\max_{\boldsymbol{z}_{i}\in\mathcal{Z}}\ \min_{\tilde{\boldsymbol{f}}_{i}\in \mathcal{F}_{2}}\mathcal{U}^{(q)}\big(\tilde{\boldsymbol{f}}_{i},\boldsymbol{z}_{i}\big)\\
		&\leq \max_{\boldsymbol{z}_{i}\in\mathcal{Z}}\ \mathcal{U}^{(q)}\big(\bar{\boldsymbol{f}},\boldsymbol{z}_{i}\big)-\min_{\tilde{\boldsymbol{f}}_{i}\in \mathcal{F}_{2}}\ \mathcal{U}^{(q)}\big(\tilde{\boldsymbol{f}}_{i},\bar{\boldsymbol{z}}\big)\\
		&\leq \max_{\boldsymbol{z}_{i}\in\mathcal{Z}}\ \mathcal{U}^{(q)}\big(\bar{\boldsymbol{f}},\boldsymbol{z}_{i}\big)-\max_{\boldsymbol{z}_{i}\in\mathcal{Z}}\ \mathcal{U}_{\mu_{\mathrm{z}}}^{(q)}\big(\bar{\boldsymbol{f}},\boldsymbol{z}_{i}\big)\leq \mu_{\mathrm{z}}D_{\mathrm{z}},
	\end{aligned} 
\end{equation}
where $D_{\mathrm{z}}\triangleq \max_{\boldsymbol{z}_{i}\in\mathcal{Z}} d(\boldsymbol{z}_{i})$. According to~\eqref{Saddle}, if $\mu_{\mathrm{z}}D_{\mathrm{z}}\leq \tilde{\epsilon}_{2}$, the solution $\bar{\boldsymbol{f}}$ satisfies~\eqref{Proximal_require}, and therefore we can set $\tilde{\boldsymbol{f}}_{i}^{(q+1)}=\bar{\boldsymbol{f}}$. However, constructing an initial primal-dual pair satisfying~\eqref{PDG_proof1} typically requires an initial value of $\mu_{\mathrm{z}}$ that is not sufficiently small to ensure $\mu_{\mathrm{z}}D_{\mathrm{z}}\leq \tilde{\epsilon}_{2}$. To address this, in each iteration, PDG reduces the value of $\mu_{\mathrm{z}}$ and then constructs a new primal-dual pair satisfying~\eqref{PDG_proof1} based on the previous one. Specifically, in the $k$th iteration, given a primal-dual pair $(\bar{\boldsymbol{f}},\bar{\boldsymbol{z}})$, the value of $\mu_{\mathrm{z}}$ is updated as $\mu_{\mathrm{z}}\leftarrow (1-\theta)\mu_{\mathrm{z}}$ with $\theta=2/(k+3)$, and then an intermediate variable $\hat{\boldsymbol{z}} = (1-\theta)\bar{\boldsymbol{z}}+\theta\bar{\boldsymbol{z}}_{i}^{(q)}(\bar{\boldsymbol{f}},\mu_{\mathrm{z}})$ is introduced to construct a new pair, expressed as $\bar{\boldsymbol{f}}\leftarrow(1-\theta)\bar{\boldsymbol{f}}+\theta\bar{\boldsymbol{f}}_{i}^{(q)}(\hat{\boldsymbol{z}})$ and $\bar{\boldsymbol{z}}\leftarrow\boldsymbol{v}_{i}^{(q)}(\hat{\boldsymbol{z}})$. Through appropriately initializing $\mu_{\mathrm{z}}$, decreasing its value in each PDG iteration, and correspondingly updating $(\bar{\boldsymbol{f}},\bar{\boldsymbol{z}})$, the condition $\mu_{\mathrm{z}}D_{\mathrm{z}}\leq \tilde{\epsilon}_{2}$ can be guaranteed after a certain number of PDG iterations~\cite{nesterovExcessiveGap2005}. 

The PP-PDG algorithm is presented in~\textbf{Algorithm~\ref{alg3}}, where the PDG algorithm is included from step 5 to step 11. The PDG iterations stop once $\hat{U}\big(\bar{\boldsymbol{f}},\tilde{\boldsymbol{f}}_{i}^{(q)}\big)-\mathcal{U}^{(q)}\big(\bar{\boldsymbol{f}}_{i}^{(q)}(\bar{\boldsymbol{z}}),\bar{\boldsymbol{z}}\big)\leq\tilde{\epsilon}_{2}$. The PP-PDG iterations stop once \eqref{Proximal_error_require} holds or $q\geq Q$, where $Q$ is the predetermined maximum number of iterations.

\begin{algorithm}[!t]
	\caption{PP-PDG-MS Scheme}\label{alg4}
	\begin{algorithmic}[1]
		\STATE \emph{Input:} $\varphi_{i}$, $\{\psi_m\}_{m\in \mathcal{I}_{\mathrm{M}}}$, $\big\{\gamma_m\big\}_{m\in \mathcal{I}_{\mathrm{M}}}$,  $\{\boldsymbol{A}_{m}\}_{m\in \mathcal{I}_{\mathrm{M}}}$, $\epsilon_{\mathrm{f}}$, $\epsilon_{\varphi}$, $\epsilon_{\mathrm{max}}$, $\epsilon_{\mathrm{min}}$, $\Delta \rho_{2}$, $\Delta_\varphi$, $Q$, $w_{\mathrm{max}}$, $w_{\mathrm{min}}$, $w_{\mu}$
		\STATE Initialize $\psi_{i+1}^{\mathrm{lb}}\leftarrow \varphi_{i}$, $\varphi_{i+1}^{(0)}\leftarrow \varphi_{i}+\Delta_\varphi$, $\Delta\varphi_{i+1}\leftarrow \Delta_\varphi$, $p\leftarrow 0$, $q\leftarrow 0$, $\rho\leftarrow 0$, $w\leftarrow w_{\mathrm{max}}$ and $\epsilon_{3}\leftarrow \epsilon_{\mathrm{max}}$. Randomly initialize $\tilde{\boldsymbol{f}}_{i}^{(0)}$ satisfying $\Vert\tilde{\boldsymbol{f}}_{i}^{(0)}\Vert_2^2=1$.
		\WHILE{$\vert\Delta\varphi_{i+1}\vert>\epsilon_{\varphi}$}
		\STATE Obtain $\tilde{\boldsymbol{f}}_{i}^{(q_\star)}$ and $q_\star$ via~\textbf{Algorithm~\ref{alg3}}. 
		\STATE Obtain $\hat{\boldsymbol{f}}_{i}^{(q_\star)}$ via~\eqref{PP_PDG_solution}. 
		\IF {$q_\star > Q$}
		\STATE $w\leftarrow \max\{w/2,w_{\mathrm{min}}\}$.
		\ELSIF {both $1-\Vert\tilde{\boldsymbol{f}}_{i}^{(q_\star)}\Vert_{2}^{2}>\epsilon_{\mathrm{f}}$ and $\vert\Delta\varphi_{i+1}\vert<\Delta_\varphi$ hold \OR \eqref{Violation1} holds}
		\STATE Update $\rho_{2} \leftarrow \rho_{2}+\Delta\rho_{2}$ and $\epsilon_{3}\leftarrow\max\{\epsilon_{3}/2,\epsilon_{\mathrm{min}}\}$.
		\ELSIF {$U(\hat{\boldsymbol{f}}_{i}^{(q_\star)})+\rho_{2}> 0$ \AND $\epsilon_{3}>\epsilon_{\mathrm{min}}$}
		\STATE Update $\epsilon_{3}\leftarrow\max\{\epsilon_{3}/2,\epsilon_{\mathrm{min}}\}$.
		\ELSE
		\IF{$U(\hat{\boldsymbol{f}}_{i}^{(q_\star)})+\rho_{2}> 0$}
		\STATE Update $\psi_{i+1}^{\mathrm{ub}}\leftarrow \varphi_{i+1}^{(p)}$, $\Delta\varphi_{i+1}\leftarrow (\varphi_{i+1}^{\mathrm{lb}}-\varphi_{i+1}^{\mathrm{ub}})/2$, and $\tilde{\boldsymbol{f}}_{i}^{(q_\star)}\leftarrow \big[\mathrm{Re}\{\hat{\boldsymbol{f}}_{i}^{\mathrm{T}}\},\mathrm{Im}\{\hat{\boldsymbol{f}}_{i}^{\mathrm{T}}\}\big]^{\mathrm{T}}$.
		\ELSE
		\STATE Update $\psi_{i+1}^{\mathrm{lb}}\leftarrow \varphi_{i+1}^{(p)}$. 
		\STATE Update $[\hat{\boldsymbol{f}}_{i}]\!\leftarrow\![\hat{\boldsymbol{f}}_{i}^{(q_\star)}]_{n}\!+\!j[\hat{\boldsymbol{f}}_{i}^{(q_\star)}]_{n+N_{\mathrm{T}}},\forall n\in\mathcal{I}_{\mathrm{N}}$.
		\IF{$\vert\Delta\varphi_{i+1}\vert<\Delta_\varphi$}
		\STATE Update $\Delta\varphi_{i+1}\leftarrow (\varphi_{i+1}^{\mathrm{ub}}-\varphi_{i+1}^{\mathrm{lb}})/2$.
		\ELSE
		\STATE Update $\epsilon_{3}\leftarrow\epsilon_{\mathrm{max}}$.
		\ENDIF
		\ENDIF
		\STATE Update $\varphi_{i+1}^{(p+1)}\leftarrow \varphi_{i+1}^{(p)}+\Delta\varphi_{i+1}$, $p\leftarrow p+1$, $w\leftarrow w_{\mathrm{max}}$ and $\rho_{2} \leftarrow 0$.
		\ENDIF	
		\STATE Update $\tilde{\boldsymbol{f}}_{i}^{(0)}\leftarrow \tilde{\boldsymbol{f}}_{i}^{(q_\star)}$.
		\ENDWHILE
		\STATE Obtain $p_\star \leftarrow p-1$. Obtain $\hat{\varphi}_{i+1}$ via~\eqref{Optimization_solution2}.
		\STATE \emph{Output:} $\hat{\boldsymbol{f}}_{i}$, $\hat{\varphi}_{i+1}$. 
	\end{algorithmic}
\end{algorithm}

\subsection{Mixed Search Method}\label{MixedSearchMethod}
In Section~\ref{Bisection_Search}, we introduced~\eqref{DC_SDR_Initialization} for initialization so that the total number of iterations of the SDR and DC algorithm can be reduced. However, the same strategy cannot be applied here since formulating a problem similar to~\eqref{DC_SDR_Initialization} for~\eqref{maxmin} is challenging. To improve the computational efficiency, we propose an adaptive MS method to replace the sole BiS. The searching step size is denoted as $\Delta \psi_{i+1}$, and the AoD in the next TS iteration is given by $\varphi_{i+1}^{(p+1)}=\varphi_{i+1}^{(p)}+\Delta\varphi_{i+1}$.

We start MS by employing a monotonic search method to efficiently establish both a tight lower bound and a tight upper bound for $\varphi_{i+1}^{\star}$. Specifically, we set $\varphi_{i}<\psi_{i+1}^{(0)}\ll \varphi_{i+1}^{\star}$ and assign $\Delta\varphi_{i+1}$ with a small positive value $\Delta_\varphi$. To streamline the process of confirming $\varphi_{i+1}^{(p)}<\varphi_{i+1}^{\star}$, at the beginning of each TS iteration, we initialize $\epsilon_{3}$ with a relatively large value, denoted as $\epsilon_{\mathrm{max}}$. Additionally, since the complexity of PP-PDG increases with the value of $L=2+2\rho_{2}$, as will be detailed in Section~\ref{complexity_PP}, we temporarily set $\rho_{2}=0$. In the $p$th TS iteration, with $\varphi_{i+1}=\varphi_{i+1}^{(p)}$, we first obtain $\tilde{\boldsymbol{f}}_{i}^{(q_\star)}$ from~\textbf{Algorithm~\ref{alg3}}, and then 
check the degree of violation for~\eqref{Constant_Constraint}. To accelerate the monotonic search stage, we adopt a relaxed criterion for violations. The validation is performed through formulating the following vector
\begin{equation}\label{PP_PDG_solution}
	\big[\hat{\boldsymbol{f}}_{i}^{(q_\star)}\big]_{n} = \frac{\big[\tilde{\boldsymbol{f}}_{i}^{(q_\star)}\big]_{n}}{\sqrt{N_{\mathrm{T}}\Big(\big[\tilde{\boldsymbol{f}}_{i}^{(q_\star)}\big]_{n}^{2}+\big[\tilde{\boldsymbol{f}}_{i}^{(q_\star)}\big]_{n+N_{\mathrm{T}}}^{2}\Big)}},
\end{equation}
where $n=1,2,\dots,2N_{\mathrm{T}}$. We update $\rho_{2} \leftarrow \rho_{2}+\Delta\rho_{2}$ and $\epsilon\leftarrow \epsilon/2$ if the following condition is met
\begin{equation}\label{Violation1}
	\big(U(\tilde{\boldsymbol{f}}_{i}^{(q_\star)})+\rho_2\Vert\tilde{\boldsymbol{f}}_{i}^{(q_\star)}\Vert_{2}\big)\big(U(\hat{\boldsymbol{f}}_{i}^{(q_\star)})+\rho_{2}\big)\leq 0.
\end{equation}
If~\eqref{Violation1} is not met, we determine the relationship between $\varphi_{i+1}^{(p)}$ and $\varphi_{i+1}^{\star}$ based on the value of $U(\hat{\boldsymbol{f}}_{i}^{(q_\star)})$. If $U(\hat{\boldsymbol{f}}_{i}^{(q_\star)})+\rho_{2}\leq 0$, a feasible solution for~\eqref{Feasibility} can be generated based on $\hat{\boldsymbol{f}}_{i}^{(q_\star)}$ as $[\hat{\boldsymbol{f}}_{i}]=[\hat{\boldsymbol{f}}_{i}^{(q_\star)}]_{n}+j[\hat{\boldsymbol{f}}_{i}^{(q_\star)}]_{n+N_{\mathrm{T}}},\ \forall n\in\mathcal{I}_{\mathrm{N}}$, and therefore we have $\varphi_{i+1}^{(p)}<\varphi_{i+1}^{\star}$. Then, to find a certain upper bound for $\psi_{i+1}^{\star}$, we adjust $\varphi_{i+1}$ upward by setting $\varphi_{i+1}^{(p+1)}\leftarrow \varphi_{i+1}^{(p)}+\Delta \varphi_{i+1}$, where $\Delta\varphi_{i+1}=\Delta_\varphi$. If $U(\hat{\boldsymbol{f}}_{i}^{(q_\star)})+\rho_{2}> 0$, the relationship between $\varphi_{i+1}^{(p)}$ and $\psi_{i+1}^{\star}$ is uncertain due to the large value of $\epsilon_{3}$. Therefore, we update $\epsilon_{3}\leftarrow\epsilon_{3}/2$. To balance the complexity and performance, we define a lower bound for $\epsilon_{3}$ as $\epsilon_{\mathrm{min}}$. If $U(\hat{\boldsymbol{f}}_{i}^{(q_\star)})+\rho_{2}\leq 0$ is observed before satisfying $\epsilon_{3} \leq \epsilon_{\mathrm{min}}$, the monotonic search continues; otherwise, we have $\varphi_{i+1}^{(p)}>\varphi_{i+1}^{\star}$ and stop the monotonic search. 

After obtaining the lower and upper bounds, we proceed with BiS and impose a higher precision requirement for PP-PDG. The procedures of BiS here are similar to those in Section~\ref{Bisection_Search}, with the only difference being the adjustment of $\rho_{2}$. We update $\rho_{2} \leftarrow \rho_{2}+\Delta\rho_{2}$ if $1-\Vert\tilde{\boldsymbol{f}}_{i}^{(q_\star)}\Vert_{2}^{2}>\epsilon_{\mathrm{f}}$, where $\epsilon_{\mathrm{f}}$ is a predetermined violation threshold.

Note that MS is computationally more efficient than BiS primarily because it requires fewer TS iterations to confirm $\varphi_{i+1}^{(p)}>\varphi_{i+1}^{\star}$. Specifically, if $\varphi_{i+1}^{(p)}\leq \varphi_{i+1}^{\star}$, PP-PDG can be stopped prematurely once $U(\hat{\boldsymbol{f}}_{i}^{(q_\star)})+\rho\leq 0$ is achieved; otherwise, the feasibility of the beamformer remains uncertain until a nearly-$\epsilon_{\mathrm{min}}$ stationary solution is found. Moreover, the computational complexity of PP-PDG increases with $M_{i}^{(p)}$ since a larger $M_{i}^{(p)}$ results in more gradient computations and typically leads to a larger $L_{i}^{(q)}$. Therefore, focusing the search within the angle interval $(\hat{\varphi}_{i},\varphi_{i+1}^{\star})$ instead of $(\hat{\varphi}_{i},+\infty)$ can enhance the computational efficiency.

\subsection{Complexity of PP-PDG-MS}\label{complexity_PP}
The proposed PP-PDG-MS scheme is summarized in~\textbf{Algorithm~\ref{alg4}}. According to~\eqref{Proxi_prove1} and~\eqref{Proxi_prove2}, the number of PP iterations required for obtaining a nearly $\epsilon_{3}$-stationary point is not larger than
\begin{equation}\label{PP_Complexity} 	
	q_{\star}\leq\big(U(\tilde{\boldsymbol{f}}_{i}^{(0)})-U(\tilde{\boldsymbol{f}}_{i}^{(q_\star+1)})\big)/\big(\epsilon_{3}^2\mu^2\sigma/8-\tilde{\epsilon}_{2}\big).
\end{equation}
The number of PDG iterations required to achieve $\mu_{\mathrm{z}}D_{\mathrm{z}}\leq \tilde{\epsilon}_{2}$ is at most $\sqrt{4\Vert\boldsymbol{G}_{i}^{(q)}\Vert_{2}^{2}D_{\mathrm{z}}/(\tilde{\epsilon}_{2}\sigma)+1/4}-3/2$~\cite{nesterovExcessiveGap2005}. Without loss of generality, we set $\mu = w_{\mu}/L$ and $\tilde{\epsilon}_{2}=w\epsilon_{3}^2\mu^2\sigma/8=ww_\mu(1-w_\mu)\epsilon_{3}^2/(8L)$, where $w_{\mu}\in(0,1)$ is a predetermined parameter, and $w\in(0,1)$ is a weight variable. The value of $w$ is initially set as $w_\mathrm{max}$, and then updated to $\max\{w/2,w_{\mathrm{min}}\}$ when $q_{\star}$ is larger than $Q$. The complexity of PP-PDG can be expressed as $\mathcal{O}\big(32\sqrt{2}LD_{\mathrm{z}}^{0.5}\tilde{w}\tilde{g}_{i}w_{\mu}^{-1}(1-w_{\mu})^{-2}\epsilon_{3}^{-3}\big)$, where $\tilde{w}=\max_{w_{\min}\leq w\leq w_{\max}}(1-w)^{-1}w^{-0.5}$ and $\tilde{g}_{i}=\max_{q=1,2,\dots,q_\star}\Vert\boldsymbol{G}_{i}^{(q)}\Vert_{2}$. Note that the formulation of~\eqref{maxmin} ensures a small value of $L$ and therefore reduces the complexity of the PP-PDG-MS scheme.

\section{Simulation Results}\label{Simulation}

\begin{table}
	\caption{Simulation Parameters\label{tab:table1}}
	\centering
	\begin{tabular}{c c c}
		\hline
		Parameter & Symbol & Value\\
		\hline
		Carrier frequency & $f_\text{c}$ & 30 GHz\\
		Path loss exponent & $\eta$ & 2\\
		Reference distance & $r_0$ & 1 m\\
		Transmit power & $P_{\mathrm{T}}$ & 40 dBm\\
		Noise power & $P_{\mathrm{N}}$ & -40 dBm\\
		Velocity & $v$ & 500 km/h\\
		Railway angle & $\alpha$ & 10\textdegree\\
		\hline
	\end{tabular}
	\vspace{-0.3cm}
\end{table}

In this section, we evaluate the performance of SDR-DC-BiS and PP-PDG-MS. The parameters for the simulations are set according to Table~\ref{tab:table1}. 

\subsection{Benchmark Schemes}
For comparisons with the proposed SDR-DC-BIS and PP-PDG-MS schemes, we provide four benchmark schemes from the existing literature. Note that in these schemes, the optimization of $\boldsymbol{f}_{i}$ is not considered, the number of switched beams is predetermined, and the optimization of $\varphi_{i+1}$ is based on the beam gain approximation, e.g., $G_{\mathrm{T}}(\theta_{i})\triangleq \pi/\theta_{i}$, where $\theta_{i}\triangleq\varphi_{i+1}-\varphi_{i}$. 
For fair comparisons, we design $\boldsymbol{f}_{i}$ to maximize the minimum RSNR within $[\varphi_{i},\varphi_{i+1})$.
The considered benchmark schemes include:

\begin{enumerate}
\item{UBW scheme~\cite{vaBeamSwitching2015}:} All beams have the same beam width, i.e., $\sin\varphi_{i+1}-\sin\varphi_{i}=(\sin\psi_{\mathrm{max}}-\sin\psi_{\mathrm{min}})/N,\ \forall i= 1,2,\dots,N$. 
\item{ESC scheme~\cite{liuAdaptiveNonUniform2023a}:} The length of railway range covered by each beam is the same, i.e., $\Vert\boldsymbol{r}(\varphi_{i+1})-\boldsymbol{r}(\varphi_{i})\Vert_{2}=\Vert\boldsymbol{r}(\psi_{\mathrm{max}})-\boldsymbol{r}(\psi_{\mathrm{min}})\Vert_{2}/N,\ \forall i= 1,2,\dots,N$.
\item{NUBW scheme with the objective of maximizing the average data rate (NUBW-M)~\cite{liuAdaptiveNonUniform2023a}:} $\{\varphi_{i}\}_{i=2}^{N}$ is obtained from solving the following optimization problem
\begin{equation}\label{Bench_NUBW_M}
	\begin{aligned}
		\max_{\{\varphi_{i}\}_{i=2}^{N}}&\quad \sum_{i=1}^{N} D(\varphi_{i},\varphi_{i+1})\\
		\mathrm{s.t.}\ &\quad \varphi_{i}<\varphi_{i+1},\ i=1,2,\dots,N,
	\end{aligned}
\end{equation}
where $\varphi_{1}=\psi_{\mathrm{min}}$, $\varphi_{N+1}=\psi_{\mathrm{max}}$, $\tilde{t}_{i}\triangleq\Vert\boldsymbol{r}(\varphi_{i})-\boldsymbol{r}(\psi_{\mathrm{min})}\Vert_{2}/v$, and $D(\varphi_{i},\varphi_{i+1})$ can be expressed as
\begin{equation}
	 D(\varphi_{i},\varphi_{i+1})\!\triangleq\!\int_{\tilde{t}_{i}}^{\tilde{t}_{i+1}}\!\ln\!\bigg(\!1 + \frac{\bar{P}_{\mathrm{T}}\pi\cos^{\eta}\big(\psi(t)+\alpha\big)}{P_{\mathrm{N}}(\varphi_{i+1}-\varphi_{i})}\!\bigg)dt.
\end{equation}
\item{NUBW scheme with the objective of stabilizing the average data rate (NUBW-S)~\cite{cuiOptimalNonuniformSteady2018}:} $\{\varphi_{i}\}_{i=2}^{N}$ is optimized to minimize the average data rate gap between adjacent beams, i.e., 
\begin{equation}\label{Bench_NUBW_S}
	\begin{aligned}
		\min_{\{\varphi_{i}\}_{i=2}^{N}}&\quad \frac{1}{N}\sum_{i=2}^{N} \bigg\vert \frac{D(\varphi_{i},\varphi_{i+1})/(\tilde{t}_{i+1}-\tilde{t}_{i})}{D(\varphi_{i-1},\varphi_{i})/(\tilde{t}_{i}-\tilde{t}_{i-1})}-1 \bigg\vert \\
		\mathrm{s.t.}\ &\quad \varphi_{i}<\varphi_{i+1},\ i=1,2,\dots,N.
	\end{aligned}
\end{equation}
\end{enumerate}

\subsection{Performance Comparisons in the Far-Field Scenario}

\begin{figure}[!t]
	\centering
	\subfigure[Beam patterns]{
		\includegraphics[width=0.99\linewidth]{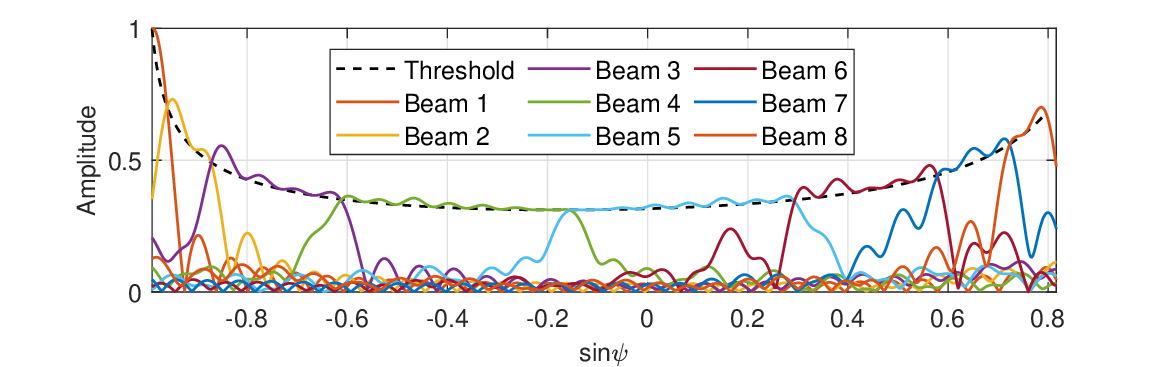}
		\label{BG_SDR}
	}\\
	\vspace{-3mm}
	\subfigure[RSNR at different HST locations]{
		\includegraphics[width=0.99\linewidth]{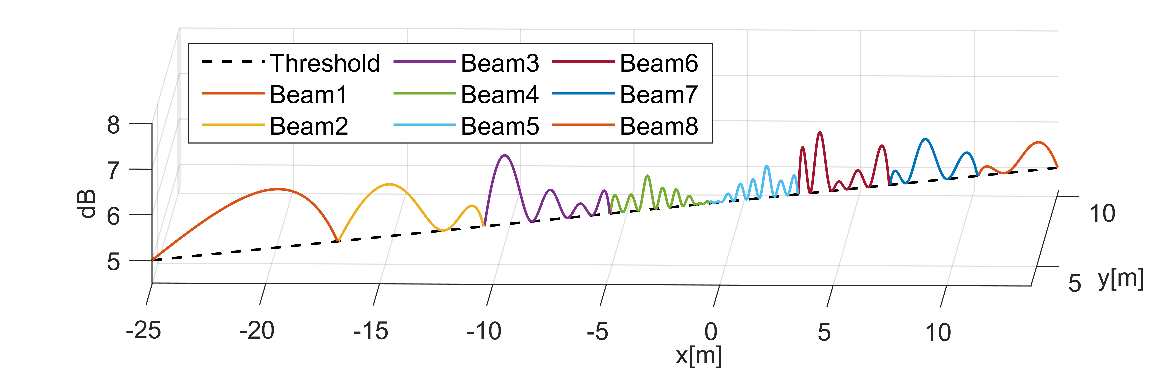}
		\label{SNR_SDR}
	}
	\vspace{-2mm}
	\caption{Designed beams using SDR-DC-BiS.}
	\vspace{-2mm}
	\label{fig:SDR_DC_BiS}
\end{figure}
\begin{figure}[!t]
	\centering
	\subfigure[Beam patterns]{
		\includegraphics[width=0.99\linewidth]{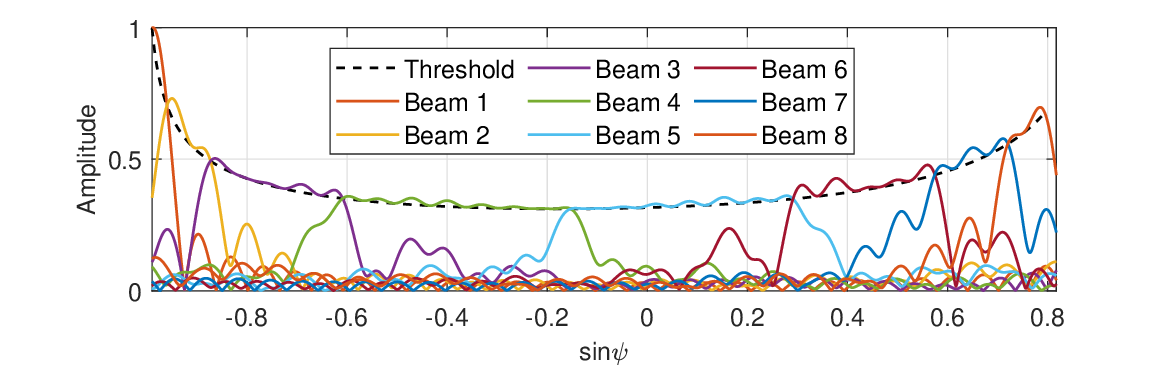}
		\label{BG_PP}
	}\\
	\vspace{-3mm}
	\subfigure[RSNR at different HST locations]{
		\includegraphics[width=0.96\linewidth]{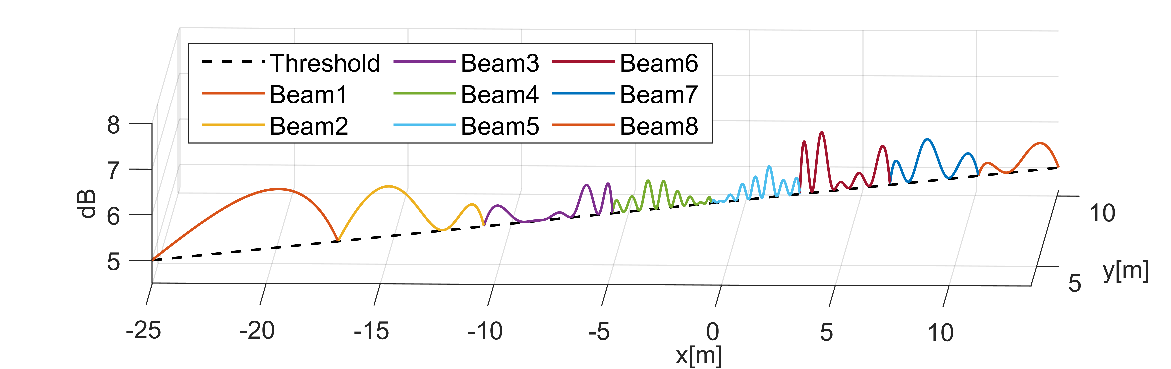}
		\label{SNR_PP}
	}
	\vspace{-2mm}
	\caption{Designed beams using PP-PDG-MS with $N_{\mathrm{T}}=32$.}
	\vspace{-2mm}
	\label{fig:PP_PDG_MS}
\end{figure}
\begin{figure}[!t]
	\centering
	\includegraphics[width=0.96\linewidth]{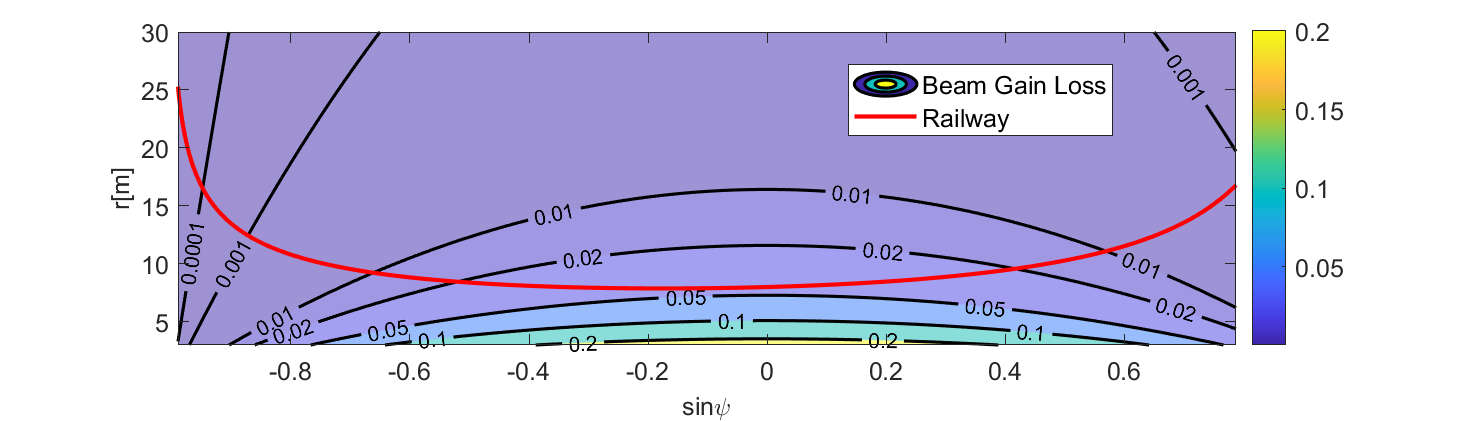}
	\vspace{-2mm}
	\caption{Value of the loss function when $N_{\mathrm{T}}=32$.}
	\vspace{-2mm}
	\label{NorF32}
\end{figure}

Given $N_{\mathrm{T}}=32$, $\gamma_{\mathrm{th}}=5$ dB, $\psi_{\mathrm{min}}=-1.4284$, $\psi_{\mathrm{max}}=0.9078$ and $y_{0}=8$ m, the designed beams using SDR-DC-BiS and PP-PDG-MS are shown in Figs.~\ref{fig:SDR_DC_BiS} and~\ref{fig:PP_PDG_MS}, respectively. The error tolerance threshold parameters are set as $\epsilon_{1}=10^{-6}$, $\epsilon_{2}=10^{-7}$, $\epsilon_{\mathrm{min}}=0.005$, $\epsilon_{\mathrm{max}}=0.05$ and $\epsilon_{\mathrm{t}}=0.005$. The weight parameters are set as $w_{\mu}=0.5$, $w_{\mathrm{max}}=0.5$ and $w_{\mathrm{min}}=0.003$. To demonstrate that the HST is in the far field of the ULA at the BS, we set the loss threshold as $\mathcal{L}_{\mathrm{th}}=0.05$ and plot the value of the loss function $\mathcal{L}(r,\psi,0,N_\mathrm{T})$ in Fig.~\ref{NorF32}. The distance from each HST location to the BS, represented by the red line, is always larger than the BAND, represented by the contour line at 0.05, indicating that the HST is always in the far-field region.

From Figs.~\ref{BG_SDR} and~\ref{BG_PP}, the beam patterns of the designed beams using SDR-DC-BiS and PP-PDG-MS exhibit negligible differences. Both schemes employ $N=8$ beams to cover the predetermined railway range. Specifically, narrow beams are employed to compensate for the high path loss at the edge of the railway range, while wide beams are used to cover the central railway range, leading to a reduced number of switched beams. Figs.~\ref{SNR_SDR} and~\ref{SNR_PP} show that the RSNR at each HST location can keep consistently at or above the given threshold. 

\begin{figure}[t]
	\centering
	\subfigure[Beam patterns]{
		\includegraphics[width=0.96\linewidth]{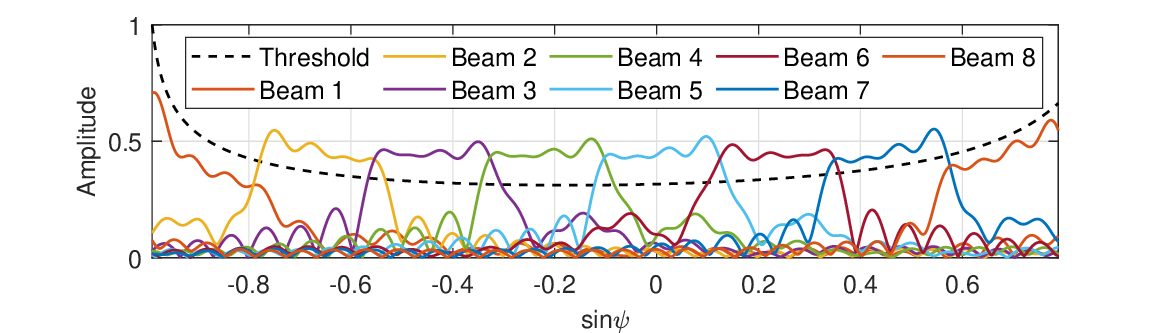}
		\label{BG_UBW}
	}\vspace{-3mm}\\
	\subfigure[RSNR at different HST locations]{
		\includegraphics[width=0.96\linewidth]{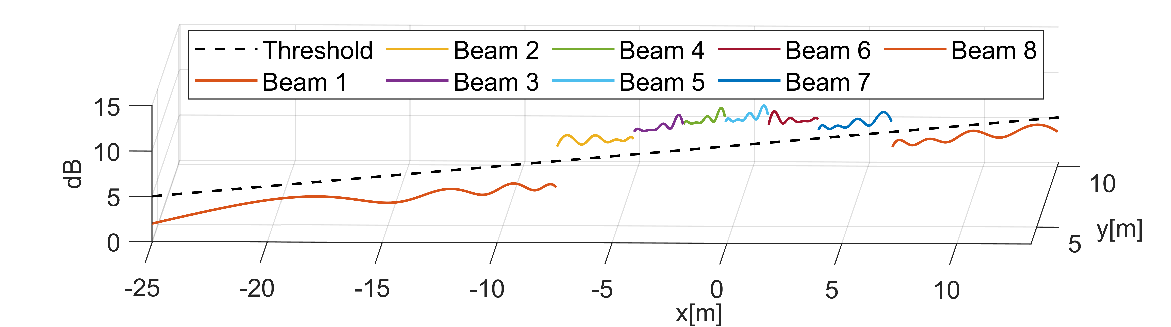}
		\label{SNR_UBW}
	}\vspace{-2mm}
	\caption{Designed beams using UBW.}
	\label{fig:UBW}
	\vspace{-2mm}
\end{figure}
\begin{figure}[t]
	\centering
	\subfigure[Beam patterns]{
		\includegraphics[width=0.96\linewidth]{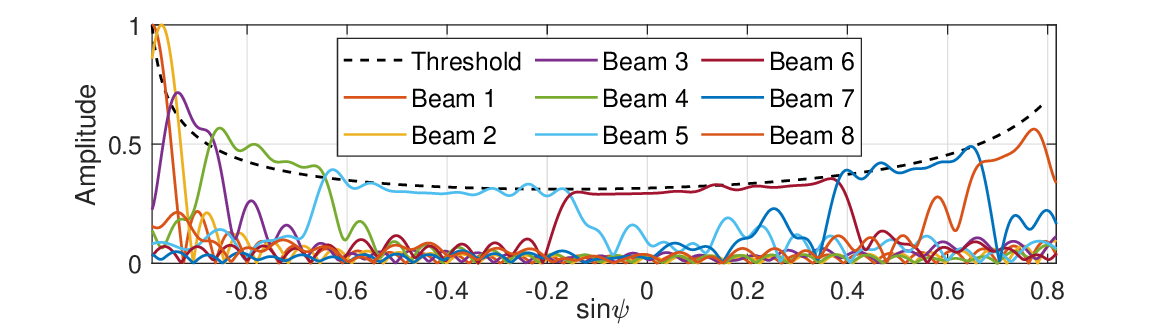}
		\label{BG_ESC}
	}\vspace{-3mm}\\
	\subfigure[RSNR at different HST locations]{
		\includegraphics[width=0.96\linewidth]{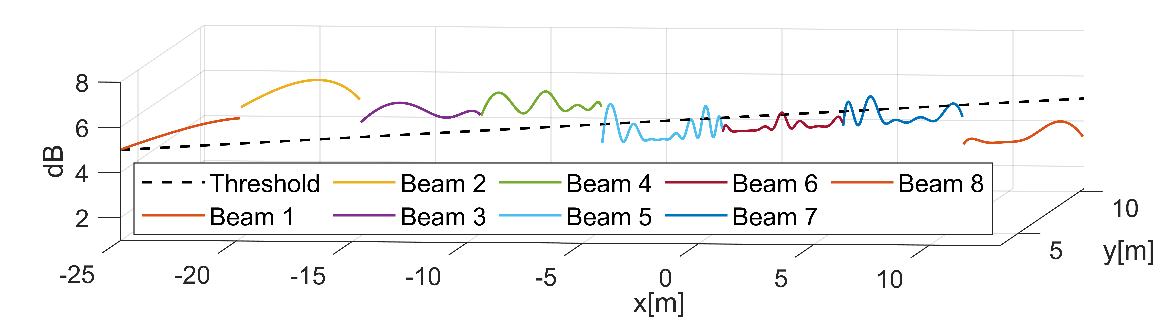}
		\label{SNR_ESC}
	}\vspace{-2mm}
	\caption{Designed beams using ESC.}
	\label{fig:ESC}
	\vspace{-2mm}
\end{figure}
\begin{figure}[t]
	\centering
	\subfigure[Beam patterns]{
		\includegraphics[width=0.96\linewidth]{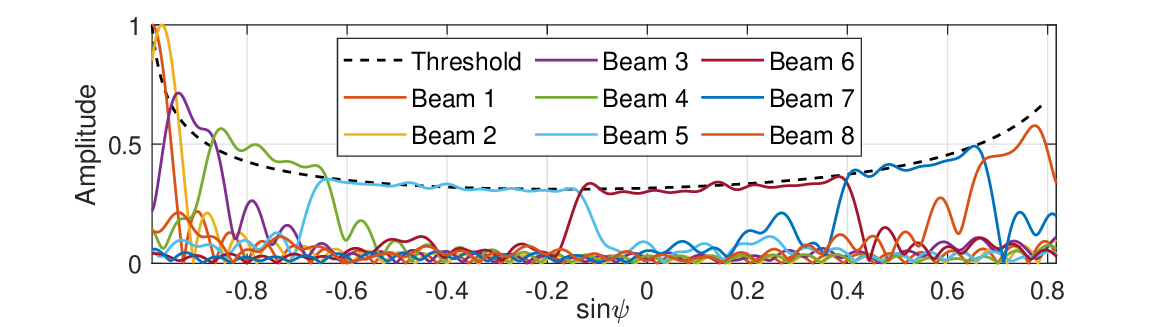}
		\label{BG_NUBWM}
	}\vspace{-3mm}\\
	\subfigure[RSNR at different HST locations]{
		\includegraphics[width=0.96\linewidth]{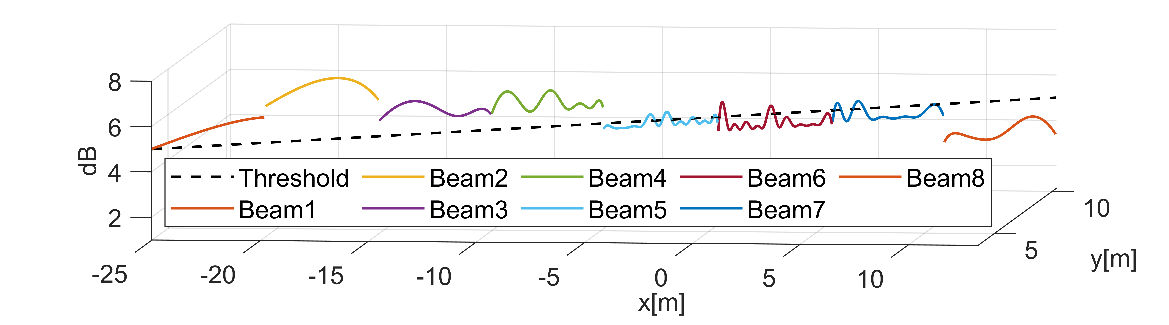}
		\label{SNR_NUBWM}
	}\vspace{-2mm}
	\caption{Designed beams using NUBW-M.}
	\label{fig:NUBW_M}
	\vspace{-2mm}
\end{figure}
\begin{figure}[t]
	\centering
	\subfigure[Beam patterns]{
		\includegraphics[width=0.96\linewidth]{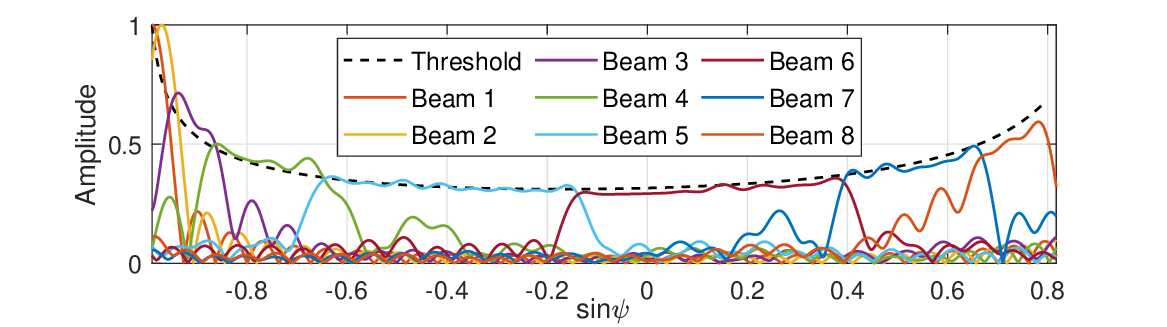}
		\label{BG_NUBWS}
	}\vspace{-3mm}\\
	\subfigure[RSNR at different HST locations]{
		\includegraphics[width=0.96\linewidth]{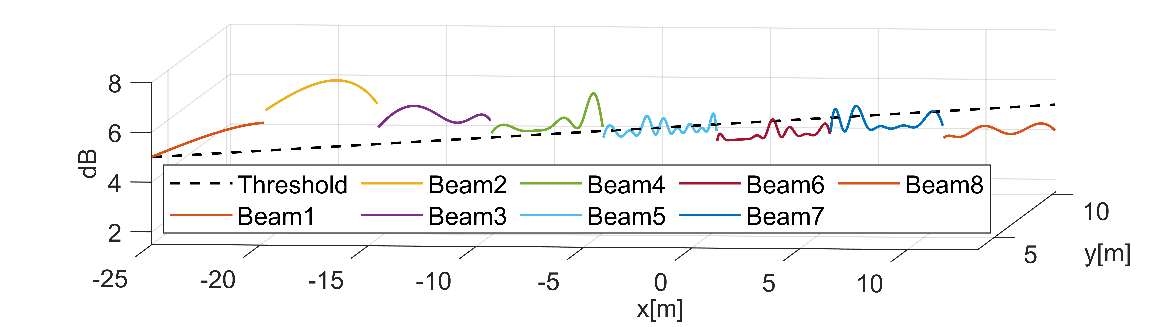}
		\label{SNR_NUBWS}
	}\vspace{-2mm}
	\caption{Designed beams using NUBW-S.}
	\label{fig:NUBW_S}
	\vspace{-2mm}
\end{figure}

The computational complexity of PP-PDG-MS is considerably lower than that of SDR-DC-BiS. To quantify the comparisons, we conduct 10 experiments for both PP-PDG-MS and SDR-DC-BiS, recording the running time for each experiment. All 20 experiments are performed using MATLAB v9.11.0.1769968 (R2021b) running on a desktop computer with a 12th Gen Intel(R) Core(TM) i9-12900K CPU at 3.20 GHz and 32GB of memory. PP-PDG-MS and  SDR-DC-BiS have average running time of 522.16 s and 13758.23 s, respectively, indicating that the running time of PP-PDG-MS is 96.20\% lower than that of SDR-DC-BiS.

The beam coverage of the designed beams is compared in Table \ref{tab:table2}. On average, the beam coverage of PP-PDG-MS is 0.0657\% narrower than that of SDR-DC-BiS, which means there is 0.0657\% performance degradation of PP-PDG-MS over SDR-DC-BiS.

\begin{table}[!t]
	\caption{Beam Coverage Results\label{tab:table2}}
	\vspace{-1mm}
	\centering
	\begin{tabular}{c c c c c c }
		\hline
		Schemes & $\hat{\varphi}_{2}$ & $\hat{\varphi}_{3}$  & $\hat{\varphi}_{5}$  & $\hat{\varphi}_{6}$ & $\hat{\varphi}_{9}$ \\
		\hline
		SDR-DC-BiS & -1.2861 & -1.0580  & -0.1537  & 0.2986 & 0.9213\\
		PP-PDG-MS &  -1.2872 &  -1.0616 &  -0.1537 &  0.2956 & 0.9179\\
		NUBW-M & -1.3548 &  -1.2410 &  -0.7057 & -0.1381 & 0.9078 \\
		NUBW-S &  -1.3551 & -1.2416 &  -0.7045 &  -0.1408 &   0.9078\\
		\hline
	\end{tabular}
	\vspace{-2mm}
\end{table}

For fair comparisons, given $N=8$, we provide the designed beams using UBW, ESC, NUBW-M, and NUBW-S in Figs.~\ref{fig:UBW},~\ref{fig:ESC},~\ref{fig:NUBW_M} and~\ref{fig:NUBW_S}, respectively. Although the six beams designed by UBW achieve higher beam gain than the predetermined threshold in Fig.~\ref{BG_UBW}, the beam gain of the other two beams is substantially lower than the threshold. Moreover, the RSNR of the eight beams varies considerably in Fig.~\ref{SNR_UBW}. In particular, the RSNR variation in the range between $[-25.02,3.59]^{\mathrm{T}}$ and $[-7.94,6.60]^{\mathrm{T}}$ is $1.99\sim3.53 $ dB, while that in the range between $[-7.94,6.60]^{\mathrm{T}}$ and $[6.21,9.10]^{\mathrm{T}}$ is $5.93 \sim 9.30$ dB. Note that 
the RSNR variation for all eight beams designed by PP-PDG-MS is only $5.00 \sim 6.31$ dB, indicating that better stability can be achieved by PP-PDG-MS than UBW. To compensate for the high path loss at the edge of the railway range, ESC, NUBW-M and NUBW-S use narrower beams than UBW. However, these three schemes allocate too many beams to cover the first half of the railway range, which results in severe instability. In particular, from Figs.~\ref{fig:ESC},~\ref{fig:NUBW_M} and~\ref{fig:NUBW_S}, the RSNR variations of ESC, NUBW-M, and NUBW-S are $3.25 \sim 7.63$ dB, $3.32 \sim7.67$ dB and $3.92\sim7.66$ dB, respectively. The essential reason of the large variation of these three schemes is the mismatch between the approximated beam gain during the optimization of the beam coverage and the actual beam gain generated from the beamformer optimization. Note that none of these three schemes can guarantee the beam gain higher than the threshold.

\subsection{Performance Evaluation in the Near-Field Scenario}

\begin{figure}[!t]
	\centering
	\includegraphics[width=0.99\linewidth]{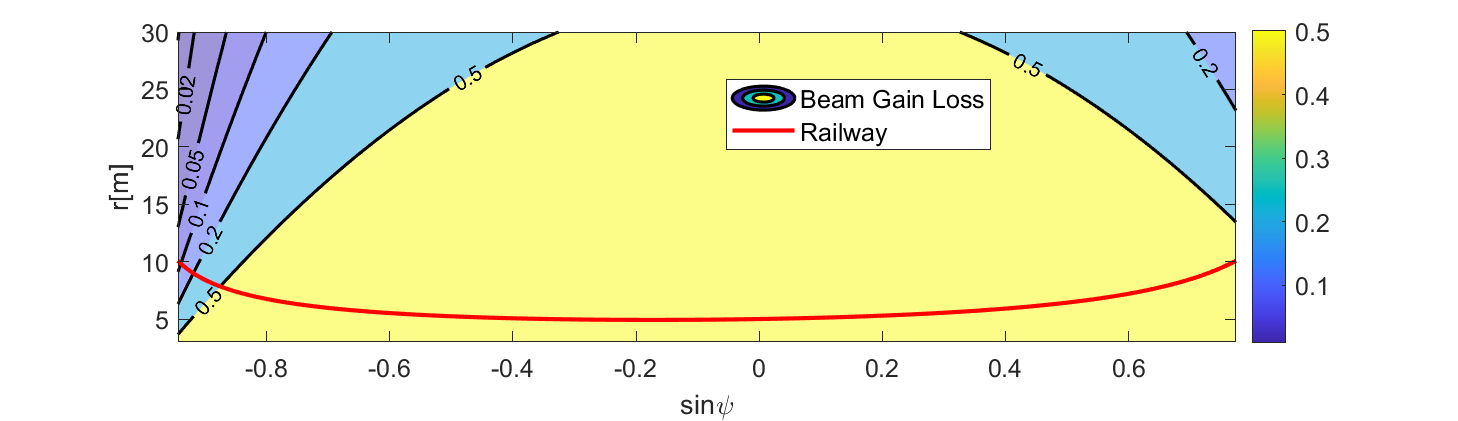}
	\caption{Value of the loss function when $N_{\mathrm{T}}=128$.}
	\label{NorF128}
\end{figure}

\begin{figure}[!t]
	\centering
	\subfigure[Beam patterns]{
		\includegraphics[width=0.975\linewidth]{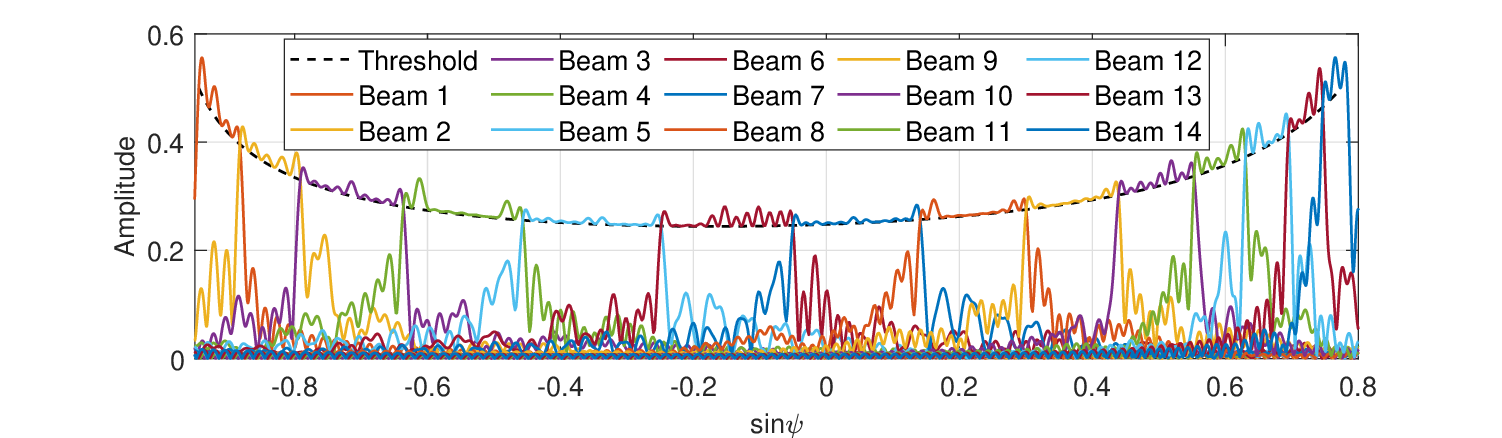}
		\label{BG_PP_128}
	}\\
	\subfigure[RSNR at different HST locations]{
		\includegraphics[width=0.99\linewidth]{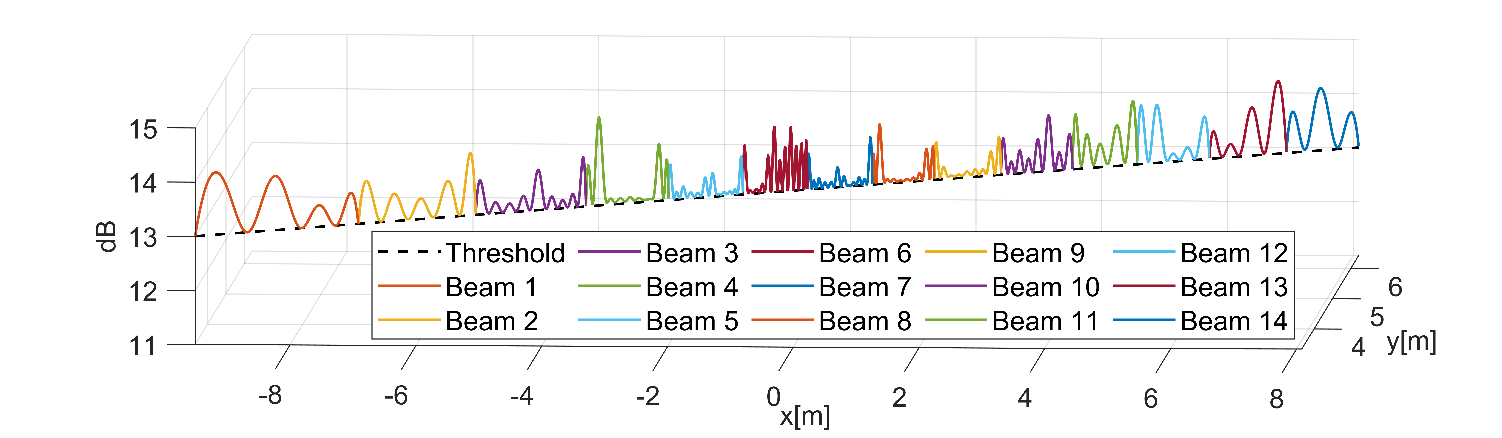}
		\label{SNR_PP_128}
	}\\
	\subfigure[Beam pattern of beam 1]{
		\includegraphics[width=0.99\linewidth]{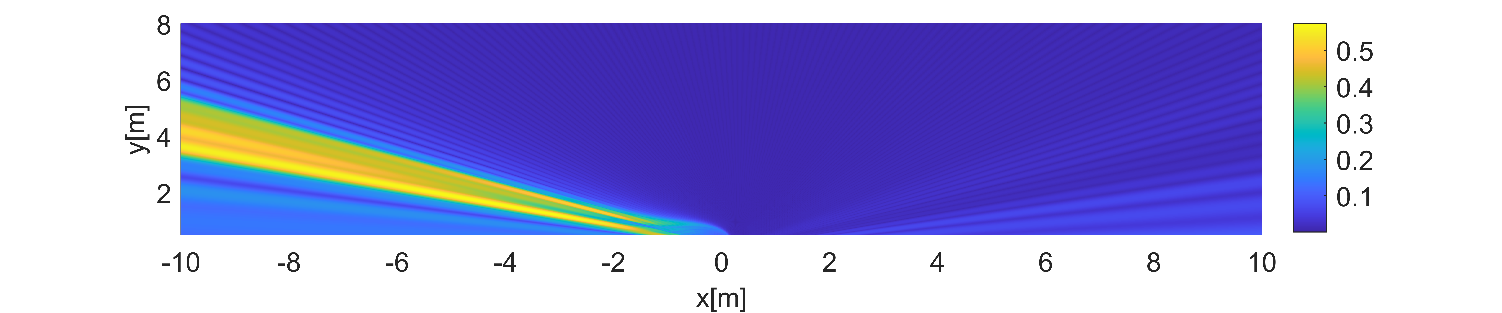}
		\label{BG_PP_Beam1}
	}\\
	\subfigure[Beam pattern of beam 8]{
		\includegraphics[width=0.99\linewidth]{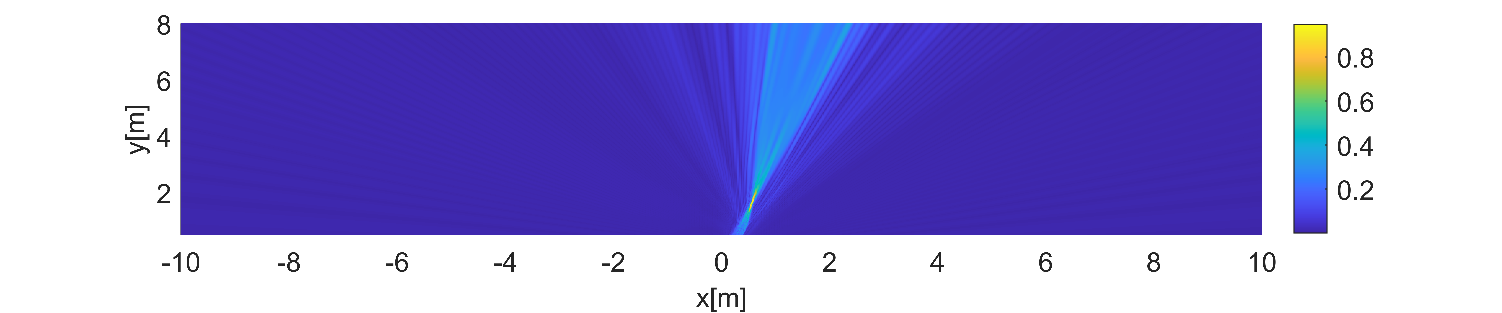}
		\label{BG_PP_Beam8}
	}\\
	\caption{Designed beams using PP-PDG-MS with $N_{\mathrm{T}}=128$.}
	\label{fig:PP_128}
\end{figure}

Since the HST may be very close to the BS, i.e., $y_0=5$ m, we also consider the near-field scenario, where we set $N_{\mathrm{T}}=128$ and $\gamma_{\mathrm{th}}=13$ dB. To demonstrate that the HST is in the near-field region, we plot the value of the loss function $\mathcal{L}(r,\psi,0,N_\mathrm{T})$ in Fig.~\ref{NorF128}, where the distance from each HST location to the BS is always smaller than the BAND, represented by the contour line at 0.05. Moreover, to evaluate the capability of PP-PDG-MS to address problems with a large number of optimization variables and RSNR constraints, we improve the sample precision to $\epsilon_{\mathrm{t}} = 0.001$ and decrease the relevant error tolerance parameters to $\epsilon_{\mathrm{min}}=0.001$ and $\epsilon_{\mathrm{max}}=0.02$. Note that SDR-DC-BiS is not suitable for this scenario due to its high computational complexity. The designed beams using PP-PDG-MS are shown in Fig.~\ref{fig:PP_128}, where $N=14$ beams are used to cover the predetermined railway range.


To illustrate the impact of the near-field effect on the designed beams, Figs.~\ref{BG_PP_Beam1} and~\ref{BG_PP_Beam8} present the patterns of beams 1 and 8 in the two-dimensional coordinate system, respectively. Since the railway range corresponding to $[\hat{\varphi}_{1},\hat{\varphi}_{2}]$ nearly falls into the far filed of the ULA at the BS, beam 1 is optimized using a series of channel samples with some far-field characteristics, and therefore exhibits the characteristics of far-field beams. As shown in Fig.~\ref{BG_PP_Beam1}, the beam gain within the mainlobe of beam 1 is generally independent of the distance. In contrast, HST locations between $\boldsymbol{r}(\hat{\varphi}_{8})$ and $\boldsymbol{r}(\hat{\varphi}_{9})$ are very close to the BS. As a result, in Fig.~\ref{BG_PP_Beam8}, beam 8 shows typical near-field characteristics, with the beam gain within its mainlobe varying with distance.

\section{Conclusion}\label{Conclusion}
In this paper, two beam design schemes, namely SDR-DC-BiS and PP-PDG-MS, have been proposed to minimize the number of switched beams within a predetermined railway range, subject to RSNR constraints and constant modulus constraints. Simulation results have demonstrated the effectiveness of both schemes and shown that compared to SDR-DC-BiS, PP-PDG-MS can achieve 96.20\% reduction in computational complexity at the cost of only 0.0657\% performance degradation. Future work will focus on efficient wide beam design.

\bibliographystyle{IEEEtran}
\bibliography{IEEEabrv.bib,IEEEexample.bib}
\vfill
	
\end{document}